\documentclass[12pt]{elsarticle}
\PassOptionsToPackage{table}{xcolor}

\makeatletter
\def\ps@pprintTitle{%
 \let\@oddhead\@empty
 \let\@evenhead\@empty
 \def\@oddfoot{\centerline{\thepage}}%
 \let\@evenfoot\@oddfoot}
\makeatother

\usepackage{hyperref}
\usepackage{amsmath,amsfonts,amssymb}
\usepackage{url}
\usepackage{bm}
\usepackage{mmacells}
\usepackage{longtable}
\usepackage[singlelinecheck=off]{caption}
\usepackage{mathtools}
\usepackage{xcolor}
\usepackage{slashed}
\usepackage{xspace}
\usepackage[absolute]{textpos}
\usepackage{braket}

\usepackage{etoolbox}
\patchcmd{\pprintMaketitle}
  {\hrule\vskip12pt}
  {\hrule\vskip12pt\ifvoid\extrainfobox\else\unvbox\extrainfobox\par\vskip12pt\fi}
  {}{}

\newsavebox\extrainfobox

\newcounter{bla}

\newcommand{\eps}{\ensuremath{\epsilon}}

\newcommand{\feyncalc}{\textsc{FeynCalc}\xspace}
\newcommand{\formcalc}{\textsc{FormCalc}\xspace}
\newcommand{\feynarts}{\textsc{FeynArts}\xspace}
\newcommand{\feynrules}{\textsc{FeynRules}\xspace}
\newcommand{\pax}{\textsc{Package-X}\xspace}
\newcommand{\ant}{\textsc{ANT}\xspace}
\newcommand{\lool}{\textsc{LOOL}\xspace}
\newcommand{\fire}{\textsc{FIRE}\xspace}

\newcommand{\ff}{\textsc{FF}\xspace}
\newcommand{\qcdloop}{\textsc{QCDLoop}\xspace}
\newcommand{\oneloop}{\textsc{OneLOop}\xspace}
\newcommand{\collier}{\textsc{Collier}\xspace}
\newcommand{\golem}{\textsc{Golem95C}\xspace}
\newcommand{\pjfry}{\textsc{PJFry}\xspace}
\newcommand{\wolfram}{\textsc{Wolfram}\xspace}
\newcommand{\gosam}{\textsc{GoSam}\xspace}
\newcommand{\fdc}{\textsc{FDC}\xspace}
\newcommand{\grace}{\textsc{GRACE}\xspace}
\newcommand{\diana}{\textsc{Diana}\xspace}
\newcommand{\feynhelpers}{\textsc{FeynHelpers}\xspace}
\newcommand{\mma}{\textsc{Mathematica}\xspace}
\newcommand{\form}{\textsc{FORM}\xspace}
\newcommand{\looptools}{\textsc{LoopTools}\xspace}
\newcommand{\tarcer}{\textsc{TARCER}\xspace}
\newcommand{\cpp}{\textsc{C++}\xspace}
\newcommand{\fare}{\textsc{FaRe}\xspace}
\newcommand{\apart}{\textsc{APart}\xspace}
\newcommand{\air}{\textsc{AIR}\xspace}
\newcommand{\litered}{\textsc{LiteRed}\xspace}

\newcommand{\reduze}{\textsc{Reduze}\xspace}
\newcommand{\nloct}{\textsc{NLOCT}\xspace}
\newcommand{\formtracer}{\textsc{FormTracer}\xspace}

\NewDocumentCommand \mmaTilde { m }
  {\ensuremath{\tilde{\text{#1}}}}

\definecolor{Gray}{gray}{0.85}
\mmaSet{morefv={gobble=2},morelst={breaklines=true}}

\begin{document}
\begin{frontmatter}

\title{FeynHelpers: Connecting FeynCalc to FIRE and Package-X}
\author[a]{Vladyslav Shtabovenko\corref{author1}}

\cortext[author1] {\textit{E-mail address:} v.shtabovenko@tum.de}

\address[a]{Technische Universit\"at M\"unchen, Physik-Department T30f, James-Franck-Str. 1, 85747 Garching, Germany}

\begin{textblock*}{30ex}(\textwidth,5ex)
TUM-EFT 75/15
\end{textblock*}

\begin{abstract}
We present a new interface called \feynhelpers that connects \feyncalc, a \mma package for symbolic semi-automatic evaluation of Feynman diagrams and calculations in quantum field theory (QFT) to \pax and \fire. The former provides a library of analytic results for scalar 1-loop integrals with up to 4 legs, while the latter is a general-purpose tool for reduction of multi-loop scalar integrals using Integration-by-Parts (IBP) identities.
\end{abstract}
\begin{keyword}
High energy physics; Feynman diagrams; Loop integrals; Dimensional regularization; Renormalization; Effective field theories; Passarino--Veltman; IBP; FeynCalc; FIRE; Package-X;

\end{keyword}

\end{frontmatter}

\section{Introduction}
\feyncalc \cite{Mertig1991,Shtabovenko2016} is firmly established in high energy physics (HEP) research and education as a flexible and easy to use tool for small and medium-sized symbolic QFT computations. Since the program by itself is mostly limited to algebraic manipulations, it is often used together with other packages, e.g.\ \feynarts, \looptools \cite{Hahn1999} or \feynrules \cite{Christensen2008}. While there is a built-in interface for converting amplitudes generated by \feynarts into valid \feyncalc input, for all the other tools one usually has to do such conversion by oneself.

As long as the other software is also written in \mma and the user has some programming experience with \wolfram language, it should be not too difficult to create a simple converter that does the job for the current project. However, there is a big difference between writing limited code for personal use, with all its potential limitations and pitfalls and creating a robust, tested and well-maintained interface that can be useful for the HEP community.

The long-term maintenance is always important because subsequent versions of a particular tool may introduce changes in syntax, changes in normalization, new features or simply workflow  improvements that require appropriate adjustments in the interface code. Also, the interface itself might require updates to be compatible with the latest releases of \mma.

In this paper we would like to present the first stable\footnote{The development version including a link to the public code repository was first mentioned in \cite{Shtabovenko2016b}.} public version of \feynhelpers, an interface that connects \feyncalc to the \mma packages \pax \cite{Patel2015} and \fire  \cite{Smirnov2013}.  The combination of \feyncalc and \feynhelpers allows one to obtain fully analytic results for most 1-loop amplitudes with up to 4 external legs and to rewrite many multi-loop amplitudes in terms of master loop integrals. Since the author of \feynhelpers is also the lead \feyncalc developer, we believe that this interface can provide a stable long-term solution for using \feyncalc, \pax and \fire in one framework, at least as long as the latter two packages are actively developed.

The main purpose for introducing \feynhelpers is not to compete with other established frameworks for loop calculations  (e.g.\ \formcalc \cite{Hahn1999}, \gosam \cite{Cullen2014}, \fdc \cite{Wang2004}, \grace \cite{Belanger2006}, \diana \cite{Tentyukov1999}) but rather to improve the usefulness of \feyncalc for specific calculations, which require semi-automatic approach and therefore cannot be easily done using fully automatic all-in-one tools.

This paper is organized as follows. Section 2 briefly introduces \mma packages \pax and \fire and explains their usefulness for automatized QFT calculations. Section 3 describes the technical implementation of the interface, which makes it possible to use both packages without interrupting the existing \feyncalc session. The installation of the interface and
its usage instructions are explained in Section 4. Section 5 presents four examples for calculations that can be significantly simplified when using \feyncalc and \feynhelpers. At the end we summarize the results and draw our conclusions in Section 6.

\section{Package-X and FIRE} \label{sec:comparison}

\subsection{Package-X}

Even though the Passarino--Veltman technique \cite{Passarino:1978jh} for evaluation of 1-loop tensor integrals was introduced almost four decades ago and alternative approaches are available (e.g.\ unitarity methods \cite{OssolaNucl.Phys.B763:147-1692007}), tensor decomposition is still widely used in many loop calculations. The key idea here is to convert all the  occurring 1-loop tensor integrals into scalar ones, which are conventionally denoted as Passarino--Veltman coefficient functions.
Thus, the calculation of an arbitrary 1-loop integral can be reduced to the evaluation of the resulting scalar functions. Often (e.g.\ in calculations of decay rates, cross sections or asymmetries) numerical evaluation of such functions is sufficient and many suitable tools for doing this are publicly available, e.g.\ \ff \cite{Oldenborgh1991}, \looptools \cite{Hahn1999}, \qcdloop \cite{Carrazza2016},
\oneloop \cite{Hameren2010}, \golem \cite{Cullen2011a}, \pjfry \cite{Fleischer2010} and \collier \cite{Denner2016}.

However, there are also cases (e.g.\ calculation of matching coefficients in effective field theories (EFTs), renormalization, etc.) where one would like to have fully analytic expressions for all the scalar functions that appear in the calculation. In general, it is desirable to have results for arbitrary kinematics, including zero Gram determinants but also for cases with vanishing and/or coinciding masses and scalar products of external momenta.

Although most of these results can be found in the literature (c.f.\ \cite{DennerNucl.Phys.B734:62-1152006} and \cite{Ellis2007}, as well as references in \cite{Patel2015}), until recently there was no easy and convenient way to make use of them in automatic calculations. Public packages such as \ant \cite{Angel2013} and \lool \cite{Ilakovac2014} aimed to provide selected results for particular kinematical limits (e.g.\ vanishing external momenta or very large masses inside loops) are available since several years, but their applicability is limited to special cases.

This situation has changed with the release of \pax \cite{Patel2015}, a \mma package for semi-automatic 1-loop calculations. A unique feature of this package is the built-in library of analytic expressions for Passarino--Veltman functions with up to 4 legs and almost arbitrary kinematics. In a tedious work the author of \pax has collected numerous analytic formulas from the literature and meticulously cross-checked everything by comparing with both analytic and numerical results. All this was then systematically implemented in an easy to use \mma package. Although \pax can do much more than just analytically evaluate coefficient functions, in this work we would like to concentrate only on this aspect of the package.

 Let us provide several examples for the usefulness of the built-in 1-loop library. When computing the 1-loop gluon self-energy in QCD (in Feynman gauge), the result depends on 3 coefficient functions:
\begin{equation}
B_0 (p^2,0,0), \quad  B_0(p^2,m^2,m^2), \quad A_0(m^2),
\end{equation}
where $p$ is the external momentum and $m$ is the mass of the quarks in the loops. Provided that \pax has already been loaded via

\begin{mmaCell}[index=1, moredefined={X}]{Input}
  <<X`
\end{mmaCell}

\begin{mmaCell}{Print}
  Package-X v2.0.1, by Hiren H. Patel
  For more information, see the guide
\end{mmaCell}
it is very easy to obtain explicit results for these coefficient functions
\begin{mmaCell}[moredefined={li, PVB, PVA}]{Input}
  li=\{PVB[0, 0, p.p, 0, 0], PVB[0, 0, p.p, m, m],
  PVA[0, m]\}
\end{mmaCell}

\begin{mmaCell}{Output}
  \{\mmaSub{B}{0}(\mmaSup{p}{2}; 0, 0), \mmaSub{B}{0}(\mmaSup{p}{2}; m, m), \mmaSub{A}{0}(m)\}
\end{mmaCell}

\begin{mmaCell}[moredefined={res, LoopRefine, li}]{Input}
  res=(LoopRefine/@li);
  res//TableForm
\end{mmaCell}
\begin{mmaCell}[addtoindex=1,form=TableForm]{Output}
  \mmaFrac{1}{\mmaTilde{\(\epsilon\)}}+log \bigg(-\mmaFrac{\mmaSup{\(\mu\)}{2}}{\mmaSup{p}{2}}\bigg)+2
  \mmaFrac{1}{\mmaTilde{\(\epsilon\)}}+\(\Lambda\)(\mmaSup{p}{2};m,m)+log\bigg(\mmaFrac{\mmaSup{\(\mu\)}{2}}{\mmaSup{m}{2}}\bigg)+2
  \mmaSup{m}{2}\bigg(\mmaFrac{1}{\mmaTilde{\(\epsilon\)}}+log\bigg(\mmaFrac{\mmaSup{\(\mu\)}{2}}{\mmaSup{m}{2}}\bigg)\bigg)+\mmaSup{m}{2}
\end{mmaCell}
where \texttt{LoopRefine} is the function that replaces coefficient functions with analytic expressions. Notice that the results returned by \pax are computed using $D = 4 - 2 \epsilon$. Furthermore, for brevity the overall factor $i/16 \pi^2$ is omitted and
$1/\eps - \gamma_E + \log (4 \pi)$ is abbreviated by $1/\tilde{\eps}$. For the same reason, the result for $B_0(p^2,m^2,m^2)$ is given as a function of $\Lambda$, which is defined as
\begin{equation}
\Lambda(p^2;m_0,m_1) = \frac{\sqrt{\lambda(p^2,m_0^2,m_1^2)}}{p^2}
\ln\Big(\frac{2 m_0 m_1}{-p^2+m_0^2+m_1^2-\sqrt{\lambda(p^2,m_0^2,m_1^2)}}+i\varepsilon\Big)\,,
\end{equation}
with
\begin{equation}
\lambda(a,b,c) = a^2 +b^2 + c^2 - 2 ab - 2ac - 2bc
\end{equation}
being the K\"{a}ll\'{e}n function. The explicit expressions can be obtained by applying \texttt{DiscExpand}

\begin{mmaCell}[moredefined={res, DiscExpand}]{Input}
  res[[2]]//DiscExpand
\end{mmaCell}

\begin{mmaCell}{Output}
  \mmaFrac{1}{\mmaTilde{\(\epsilon\)}}+\mmaFrac{\mmaSqrt{\mmaSup{p}{2}(\mmaSup{p}{2}-4\mmaSup{m}{2})}log\bigg(\mmaFrac{\mmaSqrt{\mmaSup{p}{2}(\mmaSup{p}{2}-4\mmaSup{m}{2})}+2 \mmaSup{m}{2}-\mmaSup{p}{2}}{2 \mmaSup{m}{2}}\bigg)}{\mmaSup{p}{2}}+log\bigg(\mmaFrac{\mmaSup{\(\mu\)}{2}}{\mmaSup{m}{2}}\bigg)+2
\end{mmaCell}
More examples and explanations can be found in the official tutorial\footnote{\url{http://packagex.hepforge.org/tutorial-2.0.0.pdf}}.

A \feyncalc user who is not interested in switching to a different package may naturally wonder about different possibilities to continue doing calculations in \feyncalc and use \pax only to obtain analytic results for the coefficient functions. One would, of course, like to automatize the whole procedure in such a way, that an easy and direct evaluation of \texttt{PaVe} functions with \pax can be achieved without interrupting the current \feyncalc session. This wish was indeed one of the original motivations for the development of \feynhelpers.

\subsection{FIRE}
\label{subsec:fire}
Let us now turn to another important task that arises in many loop calculations, especially when one goes beyond 1-loop. It is well known that loop integrals with propagators that are raised to integer powers usually can be reduced to simpler ones by using integration-by-parts identities (IBP) \cite{Chetyrkin1981}. The underlying formula that generates IBP relations is just a consequence of the divergence theorem in $D$-dimensional spaces
\begin{equation}
\int \frac{d^{D} {k_1}}{(2\pi)^{D}} \ldots \int \frac{d^{D} {k_n}}{(2\pi)^{D}} \frac{\partial}{\partial k^{\mu}_j}  \left (r^\mu \prod_i \frac{(p\cdot k)^{b_i}_i}{(q_i^2+m_i^2)^{a_i}} \right ) = 0, \quad \quad a_i,  b_i \in \mathbb{Z}_0 \label{eq:ibp}
\end{equation}
where $q_i$ is a linear combination of loop and external momenta (with at least one loop momentum), $(p\cdot k)_i$ denote scalar products of a loop momentum with another loop or external momentum, $r^\mu$ can be a loop or an external momentum and $j$ can take any values between $1$ and $n$. Finding a useful way to combine numerous IBP relations generated by Eq.\ \eqref{eq:ibp} such, that many complicated integrals can be reduced to a small set of master integrals, is a nontrivial task that requires special algorithms like
Laporta algorithm \cite{LaportInt.J.Mod.Phys.A15:5087-51592000}, Baikov's method \cite{BaikovPhys.Lett.B385:404-4101996} or S-bases \cite{SmirnovNucl.Phys.Proc.Suppl.160:80-842006} to name just a few of them. Often a combination of several algorithms is used to obtain the most efficient reduction for different families of loop integrals.

Several public packages (e.g.\ \air \cite{Anastasiou2004},  \litered \cite{Lee2012}, \fire \cite{Smirnov2013}, \reduze \cite{Studerus2009}) for doing IBP-reduction are available, but in this work we would like to focus specifically on \fire. It is a specialist package that was developed for doing IBP-reduction in a very general and efficient way. \fire comes with various options for fine-tuning the reduction procedure and extra utilities for using \texttt{Fermat}\footnote{\url{http://www.bway.net/lewis}} computer algebra system to speed up the calculations, and \texttt{KyotoCabinet}\footnote{\url{http://fallabs.com/kyotocabinet}} to organize the integrals in a database. Furthermore, since version 5 of the package, one has a choice between doing reduction only by means of \mma or via the (much faster) C++ back-end. In the following we would like to provide several simple examples for using \fire 5.2 with 1- and 2-loop integrals (c.f.\ \cite{Smirnov2014} for the official manual).

Let us begin with the almost trivial case of $\int  \frac{d^D q}{(q^2-m^2)^\alpha}$. The computation proceeds in 2 steps, where we first need to prepare \textit{start} files that encode the information about the given topology

\begin{mmaCell}[index=1, moredefined={FIRE5}]{Input}
  <<FIRE5`
  SetDirectory[NotebookDirectory[]];
\end{mmaCell}

\begin{mmaCell}{Print}
  FIRE, version 5.2
  DatabaseUsage: 0
  UsingFermat: False
\end{mmaCell}

\begin{mmaCell}[moredefined={Internal, External, Propagators, PrepareIBP, Prepare, SaveStart, AutoDetectRestrictions}]{Input}
  Internal= \{q\};
  External=\{\};
  Propagators= \{q^2-m^2\};
  PrepareIBP[];
  Prepare[AutoDetectRestrictions\(\pmb{\to}\)True];
  SaveStart["tadpole"];
\end{mmaCell}

\begin{mmaCell}{Print}
  Prepared
  Dimension set to 1
  No symmetries
  Saving initial data
\end{mmaCell}

\begin{mmaCell}[]{Input}
  Quit[];
\end{mmaCell}
Notice that according to the official manual, after doing so we need to restart \mma kernel, which is why the command \texttt{Quit[]} was added to the end of the code snippet. The information about the resulting integral family was saved to the file \texttt{tadpole.start}. In the next step we can perform the reduction by starting \fire again, loading \texttt{tadpole.start} and supplying explicit integer numbers for $\alpha$

\begin{mmaCell}[index=2,moredefined={LoadStart, Burn}]{Input}
  LoadStart["tadpole", 1];
  Burn[]
\end{mmaCell}

\begin{mmaCell}{Print}
  Initial data loaded
\end{mmaCell}

\begin{mmaCell}[]{Output}
  True
\end{mmaCell}

\begin{mmaCell}[moredefined={F}]{Input}
  \{F[1, \{2\}], F[1, \{3\}]\}
\end{mmaCell}

\begin{mmaCell}{Print}
  EVALUATING \{1,\{2\}\}
  Working in sector 1/1: \{1,\{1\}\}
  LAPORTA STARTED: 1 integrals for evaluation
  Maximal levels: (10)
  \{1,1\}
  Preparing points, symmetries and 2 IBP's: 0.005024`4.152594624225113 seconds.
  2 new relations produced: 0.004835`4.135941471914991 seconds.
  IRREDUCIBLE INTEGRAL: \{1,\{1\}\}
  Sector complete
  SORTING THE LIST OF 2 INTEGRALS: 0.000035`1.9956130378462462 seconds.
  Substituting 2 values: 0.000124`2.544966678658208 seconds.
  Total time: 0.022635`4.806325492327244 seconds
  Working in sector 1/1: \{1,\{1\}\}
  Sector complete
  SORTING THE LIST OF 2 INTEGRALS: 0.000038`2.031328590112781 seconds.
  Substituting 2 values: 0.000121`2.534330363812423 seconds.
  Total time: 0.002505`3.850352723699239 seconds
\end{mmaCell}

\begin{mmaCell}{Output}
  \{\mmaFrac{(d-2)G(1,\{1\})}{2\mmaSup{m}{2}},\mmaFrac{(d-4)(d-2)G(1,\{1\})}{8\mmaSup{m}{4}}\}
\end{mmaCell}
Here the first argument of $G$ identifies the integral family (the same integer was also used as the second argument of \texttt{LoadStart}), while the numbers in the brackets denote inverse powers of the original propagators. Hence, $G(1,\{1\})$  stands for $\int  \frac{d^D q}{q^2-m^2}$.

Next, let us consider the integral $\int \frac{d^D q_1 d^D q_2  \, (p \cdot q_2)^2}{q_1^2 [q_2^2]^2 [(q_1-p)^2]^3 [(q_1 - q_2 - p)^2]^3}$  which appears e.g.\ in the computation of the ghost self-energy at 2-loops in pure QCD.
With the appropriate input for \texttt{Internal}, \texttt{External} and \texttt{Propagators}
\begin{mmaCell}[index=3,moredefined={PrepareIBP, Prepare, SaveStart,Internal, External, Propagators, AutoDetectRestrictions}]{Input}
  Internal=\{q1, q2\};
  External=\{p\};
  Propagators=\{q1^2, (-p + q1)^2, (-p + q1 - q2)^2, q2^2, p q2\};
  PrepareIBP[];
  Prepare[AutoDetectRestrictions\(\pmb{\to}\)True];
  SaveStart["se2loop"];
\end{mmaCell}

\begin{mmaCell}{Print}
  Prepared
  Dimension set to 5
  No symmetries
  Saving initial data
\end{mmaCell}
the result of the reduction reads

\begin{mmaCell}[index=6]{Output}
  -\mmaFrac{(3d-10)(3d-8)(\mmaSup{d}{2}-12d+38)G(1,\{1,0,1,1,0\})}{(d-10)(d-8)(d-6)\mmaSup{p}{4}}
\end{mmaCell}

In the above examples the reduction was carried out within \mma in a reasonably small amount of time. For large calculations involving thousands or even millions of complicated multi-loop integrals, the performance of \mma is usually not sufficient such that the use of the \cpp engine becomes mandatory. However, in this work we do not want to go into details of using \fire in large scale higher loop calculations.
The reason for this is that people interested in such projects mostly use in-house \form \cite{Vermaseren2007} codes and would hardly consider \feyncalc to be appropriate for the task.

On the other hand, small and medium-sized loop calculations that are feasible with \feyncalc often may involve integrals that can be further simplified by using IBP-relations. Since \feyncalc can perform such reduction only for 2-loop propagator-type integrals (via the \tarcer \cite{Mertig1998} sub-package), the idea to utilize \fire as a general IBP-reduction back-end for \feyncalc appears tempting.

\section{Implementation}

Our goal is to be able to use \pax and \fire from an existing \feyncalc session without worrying about different syntax, conventions, and normalizations. The interface should be seamless and easy to use, while still providing access to advanced configuration options. It is not intended to provide a pass-through for every possible function present in \pax and \fire, but rather to concentrate on those functions that appear to be most useful for \feyncalc users. For example, in case of \pax we completely ignore the Lorentz and Dirac algebra modules of the package, as similar functionality is already present in \feyncalc.

First observation to make is that we cannot naively load \pax, \fire and \feyncalc in the same \mma kernel. As the symbols \texttt{DiracMatrix} and \texttt{Contract} are present both in \feyncalc and \pax, one quickly runs into shadowing issues. It is also not possible to patch \pax and change the conflicting names (as done with \feynarts), because the  package is closed-source. Fortunately, it is possible to load just the 1-loop library of \pax which does not contain any conflicting symbol names and was tested to safely coexist with \feyncalc on the same \mma kernel\footnote{The author is grateful to Hiren Patel, the developer of \pax, for explaining how to load only the \texttt{OneLoop.m} part of the package and for accounting for the compatibility with \feynhelpers while developing \pax 2.0.}. On the other hand, the internal design of \fire which puts many of the symbols introduced by the package into the \texttt{Global`} context of \mma, makes it rather impractical to use it together with \feyncalc within one kernel session. Even if that would be feasible, the necessity to quit the kernel before starting an IBP-reduction (c.f.\ Sec.\ \ref{subsec:fire}) renders the whole idea of a seamless interface meaningless. Here one can take advantage of \mma's parallel architecture that permits us to evaluate different parts of the calculation on separate kernels. This way it is possible to execute \fire on a parallel kernel and safely communicate with it using the kernel that runs \feyncalc and \feynhelpers. Before a parallel kernel is started, for each loop integral the interface creates three files: \texttt{FIREp1-intXXX.m}, \texttt{FIREp2-intXXX.m} and \texttt{FIRERepList-intXXX.m}, where \texttt{XXX} denotes an integer number assigned to that integral. The first two files are required to run \fire according to the instructions given in Sec. \ref{subsec:fire}. They can be also evaluated outside of \feynhelpers, i.e.\ directly with \mma and \fire.  The third file contains replacement rules that are used to convert the result back into \feyncalc notation. The files are located in \texttt{FeynCalc/Database} and can be examined for logging or debugging purposes.  While this approach may appear too complicated and does induce time penalties for starting and stopping parallel kernels, it allows us to avoid unnecessary restarts of the main kernel, effectively prevents variable shadowing and does not require any modifications in the source code of \fire.

Another subtlety to be taken into account when calling \pax from \feyncalc is the different normalization of 1-loop integrals. In Table \ref{tab:loopints}, we summarize the existing ways to enter a 1-loop integral in \feyncalc and \pax.

\renewcommand{\arraystretch}{1.5}
\begin{longtable}[b]{|l|l|}
\hline
Command in \feyncalc & Meaning \\
\hline
 \texttt{FAD[\{q,m1\},\{q-p,m2\}]} & $\int d^D q \, \frac{1}{[q^2-m1^2][(q-p)^2-m2^2]}$ \\
 \texttt{PaVe[0,\{SPD[p,p]\},\{$\texttt{m1}^2$,$\texttt{m2}^2$\}}] & $\int \frac{d^D q}{i \pi^2} \, \frac{1}{[q^2-m1^2][(q-p)^2-m2^2]}$ \\
  \texttt{B0[SPD[p,p],$\texttt{m1}^2$,$\texttt{m2}^2$]} \footnote{\texttt{PaVe} is the standard way to enter coefficient functions in \feyncalc. For historical reasons
it is also possible to enter several functions directly, namely via \texttt{A0}, \texttt{A00}, \texttt{B0}, \texttt{B1}, \texttt{B00}, \texttt{B11}, \texttt{C0} and \texttt{D0}.} & $\int \frac{d^D q}{i \pi^2} \, \frac{1}{[q^2-m1^2][(q-p)^2-m2^2]}$ \\
  \hline
 Command in \pax & Meaning \\
 \hline
\texttt{LoopIntegrate[1,q,\{q,m1\},\{q-p,m2\}]} & $ \frac{(4 \pi)^{D/2}}{i e^{- \gamma_E \eps}}  \int \frac{d^D q}{(2\pi)^D} \, \frac{1}{[q^2-m1^2][(q-p)^2-m2^2]}$ \\
\texttt{PVB[0, 0, p.p, m1, m2]} & $\frac{(4 \pi)^{D/2}}{i e^{- \gamma_E \eps}} \int \frac{d^D q}{(2\pi)^D} \, \frac{1}{[q^2-m1^2][(q-p)^2-m2^2]}$ \\
\hline
\caption{Different ways of entering 1-loop integrals in \feyncalc and \pax.}
\label{tab:loopints}
\end{longtable}
\pax multiplies all loop integrals by $\left (\frac{i e^{- \gamma_E \eps}}{(4 \pi)^{D/2}} \right )^{-1}$ to conveniently remove the overall $i/(16 \pi^2)$ prefactor as well as terms with $\gamma_E$ and $\log(4 \pi)$ that accompany poles in \eps. To convert a coefficient function from \feyncalc to \pax and obtain the full result, we therefore need to multiply every \texttt{PaVe}-object by
\begin{equation}
\left ( \frac{i}{16 \pi^2} \right )^{-1} \frac{i \pi^2}{(2 \pi)^D} \overset{D=4-2\eps}{=} (2 \pi)^{- 2 \eps}
\end{equation}
and perform the substitutions
\begin{align}
\frac{1}{\eps} &\to \frac{1}{\eps} - \gamma_E + \log ( 4\pi ), \\
\frac{1}{\eps^2} &\to \frac{1}{\eps^2} + \frac{1}{\eps}( - \gamma_E + \log ( 4\pi )) + \frac{\gamma_E^2}{2} - \gamma_E \log (4 \pi) + \frac{1}{2} \log^2 (4 \pi).
\end{align}
in the final result. All these steps are automatically handled by \feynhelpers, so that the user does not have to worry about different conventions. Let us also remark that in practical calculations with \feyncalc it is often convenient to omit the $1/(2\pi)^D$ prefactor in front of 1-loop integrals. The prefactor is of course understood but for brevity not written down explicitly. \feynhelpers allows us to reintroduce the prefactor when evaluating loop integrals by using the option \texttt{PaXImplicitPrefactor} (c.f.\ Sec.\ \ref{sec:examples}).

When interfacing with \fire we need to keep in mind that as of version 5.2, the package is not capable to perform reduction of integrals with linearly dependent propagators or propagators that do not form a basis. This should not be regarded as a flaw of \fire, since both decomposition into integrals with linearly independent propagators and completion of the propagator basis are operations that can be done in many ways. The choice often depends on the details of the calculation and especially on the preferred basis of master integrals.

In Sec.\ 3.3 of \cite{Shtabovenko2016} it was shown how one can handle these issues in \feyncalc by using the functions \texttt{ApartFF}, \texttt{FC\-Loop\-Basis\-FindCompletion}, \texttt{FC\-Loop\-Basis\-IncompleteQ} and \texttt{FC\-Loop\-Basis\-OverdeterminedQ}. The interface takes advantage of this new functionality by always checking the completeness of the propagator basis before passing the integrals to \fire. Automatic completion is performed when necessary, although it is also possible to supply a list of extra propagators by hand. For integrals with linearly dependent propagators the user receives a message to perform partial fractioning with \texttt{ApartFF} first.

Finally, let us briefly mention the existing alternatives to \feynhelpers. As far as \pax is concerned, we are not aware of any other publicly available interface between this package and \feyncalc. For \fire several attempts to facilitate the transition from \feyncalc exist. The \apart\footnote{\url{https://github.com/F-Feng/APart}} \cite{Feng2012a,Feng2016} package can do both partial fractioning (\texttt{ApartAll}) and basis completion (\texttt{ApartComplete}). It is also possible to convert loop integrals from \feyncalc into the \fire notation via \texttt{FireArguments}. Another package, called \fare\footnote{\url{https://sourceforge.net/projects/feyntoolfare}} \cite{Fiorentin2015},
is intended for tensor reduction of loop integrals. The resulting scalar integrals can be converted to the \fire notation via \texttt{FIREType}. It is worth noting, however, that neither \apart nor \fare currently make it possible to automatize the whole workflow, which includes not only converting the integrals but also preparing input files, running \fire, fetching the results and sending them back to \feyncalc, as it is done in \feynhelpers.

\section{Installation and Usage}

\subsection{Installation}
\feynhelpers is implemented as an add-on for \feyncalc 9.2 and above and is licensed under the General Public License (GPL) version 3. We would like to stress that this license applies only to the interface itself, but not to \pax and \fire. Their licensing conditions are outlined on the corresponding websites\footnote{\parbox[t]{10cm}{\url{http://packagex.hepforge.org} \\ \url{http://science.sander.su/FIRE.htm}}}. Questions with regard to the usage of \feynhelpers can be posted to the \feyncalc mailing list\footnote{\url{https://feyncalc.github.io/forum}}. To install \feynhelpers using the online installer, it is sufficient to evaluate

\begin{mmaCell}[index=1,moredefined={InstallFeynHelpers}]{Input}
  Import["https://raw.githubusercontent.com/FeynCalc/feynhelpers/master/install.m"]
  InstallFeynHelpers[]
\end{mmaCell}
The installer automatically suggests downloading and installing \pax and \fire if the packages are not found. To use the add-on, one should specify that it should be loaded before starting \feyncalc.
\begin{mmaCell}[index=1,moredefined={$LoadAddOns,FeynCalc}]{Input}
  $LoadAddOns=\{"FeynHelpers"\};
  << FeynCalc`
\end{mmaCell}
Essentially there are only two functions that handle all the communication between \feyncalc and the two other packages: \texttt{PaXEvaluate} and \texttt{FIREBurn}.

\subsection{\texttt{PaXEvaluate}}

\texttt{PaXEvaluate} is the main function of the \pax-interface. It works on scalar loop integrals without any loop-momentum dependent scalar products in the numerator and on Passarino--Veltman functions. If one wants to obtain only the UV- or the IR-divergent part of the result, one can use \texttt{PaXEvaluateUV} and \texttt{PaXEvaluateIR}. Finally, \texttt{PaXEvaluateUVIRSplit} returns the full result with an explicit distinction between $\eps_\textrm{UV}$ and $\eps_\textrm{IR}$. All four functions share the same set of options.

\begin{quote}
\renewcommand{\arraystretch}{2}
\rowcolors{1}{Gray}{Gray}
\setlength{\tabcolsep}{0.6cm}
\begin{longtable}[b]{|l p{5.8cm}|}
\hline
\texttt{PaXEvaluate}[\textit{expr}, \textit{q}] &  converts all the scalar 1\,-loop integrals with loop momentum $q$ in \textit{expr} to Passarino--Veltman functions, which are then analytically evaluated using \pax. Both UV- and IR-singularities are regulated with \eps. If  $q$ is omitted, no conversion is done and only already present Passarino--Veltman functions are evaluated.  \\[0.2cm]
\texttt{PaXEvaluateUV}[\textit{expr}, \textit{q}] &  like \texttt{PaXEvaluate}, but only the $1/{\eps_\textrm{UV}}$-piece of the result is returned. \\[0.2cm]
\texttt{PaXEvaluateIR}[\textit{expr}, \textit{q}] &  like \texttt{PaXEvaluate}, but only the $1/{\eps_\textrm{IR}}$-piece of the result is returned. \\[0.2cm]
\texttt{PaXEvaluateUVIRSplit}[\textit{expr}, \textit{q}] &  like \texttt{PaXEvaluate}, but  with the explicit distinction between $1/{\eps_\textrm{UV}}$ and $1/{\eps_\textrm{IR}}$ in the final result. \\[0.2cm]
  \hline
\end{longtable}
\end{quote}

\begin{quote}
\renewcommand{\arraystretch}{2}
\rowcolors{1}{Gray}{Gray}
\begin{longtable}{|l  p{3cm} p{5.5cm}|}
    \hline
    \textbf{Option} & \textbf{Default value} & \textbf{Description} \\
    \hline
\texttt{Collect} & \texttt{True} & whether the result should be collected with respect to scalar products, metric tensors and the Levi-Civita tensors. \\
  \texttt{Dimension} & \texttt{D} & the symbol that denotes $D$-dimensions in the loop integrals.  \\
    \texttt{FCE} & \texttt{False} & whether the result should be converted into \texttt{FeynCalcExternal}-notation. \\
\texttt{FCVerbose} & \texttt{False} & allows us to activate the debugging output. \\
  \texttt{FinalSubstitutions} & \texttt{\{\}} & list of replacements to be applied to the final result. \\
      \texttt{PaVeAutoOrder} & \texttt{True} & automatic ordering of arguments inside FeynCalc's \texttt{PaVe} functions. \\
  \texttt{PaVeAutoReduce} & \texttt{True} & automatic reduction of certain \texttt{PaVe} functions into simpler ones. \\
    \texttt{PaXC0Expand} & \texttt{False} & whether the full analytic result for the $C_0$ function should be inserted. \\
  \texttt{PaXD0Expand} & \texttt{False} & whether the full analytic result for the $D_0$ function should be inserted. \\
  \texttt{PaXDiscExpand} & \texttt{True} & whether the \texttt{DiscB} function of \pax should be replaced with its explicit expression. \\
  \texttt{PaXExpandInEpsilon} & \texttt{True} &  whether the final results multiplied by \texttt{PaXImplicitPrefactor} should be expanded around $4 - 2 \eps$. \\
  \texttt{PaXImplicitPrefactor} & \texttt{1} & a $D$-dependent prefactor that multiplies the final result. \\
  \texttt{PaXKallenExpand} & \texttt{True} & whether the $\texttt{Kallen}\lambda$ function of \pax should be replaced with its explicit expression. \\
  \texttt{PaXLoopRefineOptions} & \{\} & allows us to directly specify options for \texttt{LoopRefine} of \pax. \\
  \texttt{PaXPath} & \texttt{FileNameJoin}[\{\allowbreak \$\texttt{UserBase\-Directory,"App\-lications}",\texttt{"X"}\}] & path to \pax. \\
  \texttt{PaXSimplifyEpsilon} & \texttt{True} & whether the finite and the divergent parts of the result should be simplified via \texttt{Simplify}. \\
  \texttt{PaXSubstituteEpsilon} & \texttt{True} & whether the result should be given with standard normalization, i.e., without the prefactor $\left ( \frac{i e^{- \gamma_E \varepsilon}}{(4 \pi)^{D/2}} \right )^{-1}$ introduced by \pax.  \\
    \texttt{PaXSeries} & \texttt{False} & offers the possibility to expand a Passarino--Veltman function around given parameters via \texttt{LoopRefineSeries} of \pax. \\
    \texttt{PaXAnalytic} & \texttt{True} & allows \texttt{LoopRefineSeries} to construct series expansions near Landau singularities by means of analytic continuation.  \\
  \hline
\end{longtable}
\end{quote}

\texttt{PaXEvaluate} is designed in such a way that the function requires only a minimal amount of user input. For example, to compute the integral $\int \frac{d^D q}{(2 \pi)^D} \frac{1}{q^2-m^2}$ it is sufficient to write
\begin{mmaCell}[moredefined={int, PaXEvaluate, FAD, PaXImplicitPrefactor}]{Input}
  int=PaXEvaluate[FAD[\{q,m\}],q,PaXImplicitPrefactor\(\pmb{\to}\)1/(2Pi)^D]
\end{mmaCell}
\begin{mmaCell}{Output}
  \mmaFrac{i\mmaSup{m}{2}}{16\mmaSup{\(\pi\)}{2}\(\varepsilon\)}-\mmaFrac{i\mmaSup{m}{2}\big(-log\big(\mmaFrac{\mmaSup{\(\mu\)}{2}}{\mmaSup{m}{2}}\big)+\(\gamma\)-1-log(4\(\pi\))\big)}{16 \mmaSup{\(\pi\)}{2}}
\end{mmaCell}
where the first argument is our integral, the second is the loop momentum and the third is the option to specify the normalization. To make the result look more compact, we can introduce the abbreviation $\Delta \equiv 1/\eps - \gamma_E + \log(4\pi)$. This can be done with the \feyncalc function \texttt{FCHideEpsilon}
\begin{mmaCell}[moredefined={int, FCHideEpsilon}]{Input}
  int//FCHideEpsilon
\end{mmaCell}
\begin{mmaCell}{Output}
  \mmaFrac{i\(\Delta\)\mmaSup{m}{2}}{16\mmaSup{\(\pi\)}{2}}+\mmaFrac{i\mmaSup{m}{2}\big(log\big(\mmaFrac{\mmaSup{\(\mu\)}{2}}{\mmaSup{m}{2}}\big)+1\big)}{16\mmaSup{\(\pi\)}{2}}
\end{mmaCell}
In practical calculations one is usually interested in the evaluation of different Passarino--Veltman functions. In this case it is not needed to specify the loop momentum, such that to compute e.g.\ $B_0 (p^2,0,m^2)$ we use
\begin{mmaCell}[moredefined={PaXEvaluate, B0, SPD}]{Input}
  PaXEvaluate[B0[SPD[p,p],0,m^2]]
\end{mmaCell}
\begin{mmaCell}{Output}
  \mmaFrac{1}{\(\varepsilon\)}+log\bigg(\mmaFrac{\mmaSup{\(\mu\)}{2}}{\(\pi\)\mmaSup{m}{2}}\bigg)-\mmaFrac{\mmaSup{m}{2}log\big(\mmaFrac{\mmaSup{m}{2}}{\mmaSup{m}{2}-\mmaSup{p}{2}}\big)}{\mmaSup{p}{2}}+log\bigg(\mmaFrac{\mmaSup{m}{2}}{\mmaSup{m}{2}-\mmaSup{p}{2}}\bigg)-\(\gamma\)+2
\end{mmaCell}
\pax can also expand coefficient functions in their parameters (masses or external momenta). To expand $B_0 (p^2,0,m^2)$ around $p^2 = m^2$ up to first order with \texttt{PaXEvaluate} we first need to assign an arbitrary symbolic value to the scalar product $p^2$, e.g.\ \texttt{pp}
\begin{mmaCell}[moredefined={SPD}]{Input}
  SPD[p,p]=pp;
\end{mmaCell}
Then we use the option \texttt{PaXSeries} to specify the expansion parameters and activate the option \texttt{PaXAnalytic} to ensure that the derivatives of loop integrals are taken appropriately
\begin{mmaCell}[moredefined={PaXEvaluate, B0, SPD, PaXSeries, PaXAnalytic}]{Input}
  PaXEvaluate[B0[SPD[p,p],0,m^2],PaXSeries\(\pmb{\to}\)\{\{pp,m^2,1\}\},PaXAnalytic\(\pmb{\to}\)True]
\end{mmaCell}
\begin{mmaCell}{Output}
  \mmaFrac{3 \mmaSup{m}{2}-pp}{2\(\varepsilon\)\mmaSup{m}{2}}-\mmaFrac{(3 \mmaSup{m}{2}-pp)\big(-log\big(\mmaFrac{\mmaSup{\(\mu\)}{2}}{\mmaSup{m}{2}}\big)+\(\gamma\)-2+log(\(\pi\))\big)}{2\mmaSup{m}{2}}
\end{mmaCell}
If we are interested only in the UV-part of this series, it is sufficient to replace \texttt{PaXEvaluate} with \texttt{PaXEvaluateUV}
\begin{mmaCell}[moredefined={PaXEvaluateUV, B0, SPD, PaXSeries, PaXAnalytic}]{Input}
  PaXEvaluateUV[B0[SPD[p,p],0,m^2],PaXSeries\(\pmb{\to}\)\{\{pp,m^2,1\}\},PaXAnalytic\(\pmb{\to}\)True]
\end{mmaCell}
\begin{mmaCell}{Output}
  \mmaFrac{1}{\mmaSub{\(\varepsilon\)}{UV}}
\end{mmaCell}
Similarly, we can also obtain the IR-part of the series
\begin{mmaCell}[moredefined={PaXEvaluateIR, B0, SPD, PaXSeries, PaXAnalytic}]{Input}
  PaXEvaluateIR[B0[SPD[p,p],0,m^2],PaXSeries\(\pmb{\to}\)\{\{pp,m^2,1\}\},PaXAnalytic\(\pmb{\to}\)True]
\end{mmaCell}
\begin{mmaCell}{Output}
  \mmaFrac{\mmaSup{m}{2}-pp}{2\mmaSup{m}{2}\mmaSub{\(\varepsilon\)}{IR}}
\end{mmaCell}
Finally, \texttt{PaXEvaluateUVIRSplit} returns the result with explicit distinction between UV and IR singularities
\begin{mmaCell}[moredefined={PaXEvaluateUVIRSplit, B0, SPD, PaXSeries, PaXAnalytic}]{Input}
  PaXEvaluateUVIRSplit[B0[SPD[p,p],0,m^2],PaXSeries\(\pmb{\to}\)\{\{pp,m^2,1\}\},PaXAnalytic\(\pmb{\to}\)True]
\end{mmaCell}

\begin{mmaCell}{Output}
  \mmaFrac{\mmaSup{m}{2}-pp}{2\mmaSup{m}{2}\mmaSub{\(\varepsilon\)}{IR}}-\mmaFrac{(3 \mmaSup{m}{2}-pp)\big(-log\big(\mmaFrac{\mmaSup{\(\mu\)}{2}}{\mmaSup{m}{2}}\big)+\(\gamma\)-2+log(\(\pi\))\big)}{2 \mmaSup{m}{2}}+\mmaFrac{1}{\mmaSub{\(\varepsilon\)}{UV}}
\end{mmaCell}

A comment is in place here. It is well known that in dimensional regularization (DR) it is possible to regularize both ultraviolet (UV) and infrared (IR) divergences with the same regulator $\eps$. To disentangle both types of divergences one can use different regulators, i.e.\ keep $\eps$ for the UV divergences and regulate the IR divergences with a fictitious mass. On the other hand, it is also possible to distinguish between UV and IR divergences in DR by using $\eps_\textrm{UV}$ and $\eps_\textrm{IR}$, as is done e.g., in \cite{Manohar1997}. Such a distinction is often useful in matching calculations for EFTs, but also in renormalization calculations. With this prescription the rule that scaleless integrals vanish in DR does not hold anymore. For example, the logarithmically divergent scaleless 1-loop integral
\begin{equation}
\int \frac{d^D l}{(2 \pi)^D} \frac{1}{l^4}  = \int \frac{d^D l}{(2 \pi)^D} \left ( \frac{1}{l^2 (l^2-m^2)} - \frac{m^2}{l^4 (l^2-m^2)} \right )
\end{equation}
cannot be set to zero, as it is proportional to $1/\eps_\textrm{UV} - 1/\eps_\textrm{IR}$.
\feyncalc 9.2 features a new global option \texttt{\$KeepLogDivergentScalelessIntegrals}, which prevents the internal functions from setting such scaleless 1-loop integrals to zero. Together with \texttt{PaXEvaluateUVIRSplit} this can be used to consistently distinguish between UV and IR singularities regulated dimensionally at 1-loop. For example, we can see that after setting
\begin{mmaCell}[moredefined={$KeepLogDivergentScalelessIntegrals}]{Input}
  $KeepLogDivergentScalelessIntegrals=True;
\end{mmaCell}
the integral $\int \frac{d^D l}{(2 \pi)^D} \frac{1}{l^4}$ does not vanish anymore but evaluates to
\begin{mmaCell}[moredefined={FCI, FAD, PaXEvaluateUVIRSplit, PaXImplicitPrefactor}]{Input}
  PaXEvaluateUVIRSplit[FAD[\{l,0,2\}],l, PaXImplicitPrefactor->1/(2Pi)^D]
\end{mmaCell}
\begin{mmaCell}{Output}
  \mmaFrac{i}{16\mmaSup{\(\pi\)}{2}\mmaSub{\(\varepsilon\)}{UV}}-\mmaFrac{i}{16\mmaSup{\(\pi\)}{2}\mmaSub{\(\varepsilon\)}{IR}}
\end{mmaCell}
Notice that this prescription is implemented in \feyncalc for 1-loop integrals only. Furthermore, to ensure that the results are consistent, one should not use \texttt{FIREBurn}, as by default \fire always sets all the scaleless integrals to zero.

\subsection{\texttt{FIREBurn}}

\texttt{FIREBurn} is the main function of the \fire-interface. It reduces scalar multi-loop integrals to simpler ones using IBP-techniques.

\begin{quote}
\renewcommand{\arraystretch}{2}
\rowcolors{1}{Gray}{Gray}
\setlength{\tabcolsep}{0.8cm}
\begin{longtable}[b]{|p{5.3cm} p{5.3cm}|}
\hline
\texttt{FIREBurn}[\textit{expr}, \textit{\{$q_1$, $q_2$,\ldots\}}, \textit{\{$p_1$, $p_2$,\ldots\}}] &  reduces all loop integrals in \textit{expr} that depend on the loop momenta $q_1$, $q_2$, \ldots and external momenta $p_1$, $p_2$,\ldots                              . \\
  \hline
\end{longtable}
\end{quote}

\begin{quote}
\renewcommand{\arraystretch}{2}
\rowcolors{1}{Gray}{Gray}
\begin{longtable}{|l  p{3cm} p{5.5cm}|}
    \hline
    \textbf{Option} & \textbf{Default value} & \textbf{Description} \\
    \hline
\texttt{Collect} & \texttt{True} & whether the result should be collected with respect to the loop integrals. \\
  \texttt{FCE} & \texttt{False} & whether the result should be converted into \texttt{FeynCalcExternal}-notation. \\
  \texttt{FCLoopIBPReducableQ} & \texttt{False} & whether \fire should try to reduce every loop integral in the expression (default) or only those that contain propagators raised to integer powers. \\
  \texttt{FCVerbose} & \texttt{False} & allows us to activate the debugging output. \\
  \texttt{FIREAddPropagators} & \texttt{Automatic} & whether extra propagators with zero powers needed to complete the basis should be added automatically or by hand. \\
  \texttt{FIREConfigFiles} & \texttt{\{ToFileName[{\allowbreak\$FeynCalc\-Directory, "Database"}, "FIREp1.m"],
            \texttt{ToFileName[{\allowbreak\$FeynCalc\-Directory, "Database"}, "FIREp2.m"],
            \texttt{ToFileName[{\allowbreak\$FeynCalc\-Directory, "Database"}, "FIRERepList.m"]}}\}} & where to save scripts for running \fire. \\
  \texttt{FIREPath} & \texttt{FileNameJoin[{\allowbreak\$UserBase\-Directory, "Applications", "FIRE5", "FIRE5.m"}]} & path to \fire. \\
    \texttt{FIRERun} & \texttt{True} & whether the reduction should be started after all the configuration files have been created (default). Otherwise, the interface would just create the files but not run \fire afterwards. \\
  \texttt{FIRESilentMode} & \texttt{True} & whether the (rather verbose) text output of \fire should be suppressed. \\

  \texttt{FIREStartFile} & \texttt{ToFileName[{\allowbreak\$FeynCalc\-Directory, "Database"}, "FIREStartFile"]} & where to save start file for the \fire engine. \\
  \texttt{Timing} & \texttt{True} & informs the user about the progress of the IBP-reduction. \\
  \hline
\end{longtable}
\end{quote}

The function requires only three arguments, which are the input expression, the list of loop momenta and the list of external momenta. For example, to IBP-reduce the 1-loop integral $\int \frac{ d^D l}{[l^2]^2 [(l-p)^2 -m^2]^2}$ we need to enter
\begin{mmaCell}[moredefined={int, FIREBurn, FAD}]{Input}
  FIREBurn[FAD[\{l,0,2\},\{l-p,m,2\}],\{l\},\{p\}]
\end{mmaCell}
\begin{mmaCell}{Output}
  \mmaFrac{(D-2)(2 D \mmaSup{m}{2}-9 \mmaSup{m}{2}-pp)}{2\mmaSup{m}{2}\mmaSup{(\mmaSup{m}{2}-pp)}{3}(\mmaSup{(l-p)}{2}-\mmaSup{m}{2})}-\mmaFrac{(D-3)(D \mmaSup{m}{2}+D pp-4\mmaSup{m}{2}-6 pp)}{\mmaSup{(\mmaSup{m}{2}-pp)}{3}\mmaSup{l}{2}.(\mmaSup{(l-p)}{2}-\mmaSup{m}{2})}
\end{mmaCell}
If the integral has no dependence on external momenta, as for example the 3-loop integral $\int  \frac{d^D q_1 d^D q_2  d^D q_3}{[q1^2-m^2]^2 [(q_1+q_3)^2-m^2] [(q_2-q_3)^2] [q_2^2]^2}$, then the list of external momenta should be left empty
\begin{mmaCell}[moredefined={FIREBurn, FAD}]{Input}
  FIREBurn[FAD[\{q1,m,2\},\{q1+q3,m\},\{q2-q3\},\{q2,0,2\}],\{q1,q2,q3\},\{\}]
\end{mmaCell}
\begin{mmaCell}{Output}
  -\mmaFrac{(D-3)(3 D-10)(3 D-8)}{16(2 D-7)\mmaSup{m}{4}(\mmaSup{q1}{2}-\mmaSup{m}{2}).\mmaSup{q2}{2}.\mmaSup{(q2-q3)}{2}.(\mmaSup{(q1+q3)}{2}-\mmaSup{m}{2})}
\end{mmaCell}
In the current implementation of the interface to \fire, each loop integral is evaluated separately, which is of course rather inefficient. Unfortunately, \feyncalc is still not able to automatically recognize multi-loop integrals that belong to the same topology, which is undoubtedly a crucial requirement to make the program more useful in multi-loop calculations. Despite this limitation, we believe that \texttt{FIREBurn} can be well employed in smaller calculations with a low number of loops, where the IBP-reduction makes it possible to arrive to simpler results.

\section{Examples}

\label{sec:examples}
So far we explained how to use \feynhelpers to evaluate single loop integrals. To demonstrate the usefulness of the interface in more realistic scenarios, we will provide four examples that make extensive use of the new functions introduced with \feynhelpers. The complete working codes for these calculations can be found in the accompanying \mma notebooks
\texttt{QED-Renormalization.nb}, \texttt{NRQCD.nb}, \texttt{HiggsDecay.nb} and \texttt{QED-TwoLoop-SelfEnergies.nb}. These codes are shipped together with \feynhelpers\footnote{
When \feynhelpers is loaded, the greeting message contains the sentence ``Have a look at the supplied examples''. Clicking on the word ``examples'' opens the directory with sample calculations.} and can be also viewed online\footnote{\url{https://github.com/FeynCalc/feynhelpers/tree/master/Examples}}. To avoid cluttering up this paper we will not copy the full code here but rather  merely explain the most important steps and present the final results.

\subsection{Renormalization of QED at 1-loop}
Our first example is the 1-loop renormalization of QED in three different schemes: minimal subtraction (MS), modified minimal subtraction ($\overline{\textrm{MS}}$) and on-shell (OS). This calculation should be familiar to every QFT practitioner and can be found in many books, e.g.\ \cite{Bohm:2001yx} which we will follow here.

We start with
\begin{align}
 \mathcal{L}_{\textrm{QED}} &=  \mathcal{L}_{\textrm{R,QED}} +  \mathcal{L}_{\textrm{CT}}, \\
  \mathcal{L}_{\textrm{R,QED}} & = -\frac{1}{4} F_{\mu \nu}F^{\mu \nu} - \frac{1}{2 \xi}(\partial^\mu A_\mu)^2 +
  \bar{\psi} (i\slashed{\partial} -m) \psi + e  \bar{\psi} \slashed{A} \psi ,\\
  \mathcal{L}_{\textrm{CT}}  & = - (Z_A-1) \frac{1}{4}F_{\mu \nu}F^{\mu \nu} - \frac{1}{2\xi} (Z_A Z^{-1}_\xi - 1) (\partial^\mu A_\mu)^2 \nonumber \\
    & + (Z_\psi-1) \bar{\psi} i\slashed{\partial} \psi- (Z_\psi Z_m -1) m \bar{\psi} \psi + (Z_\psi Z_A^{1/2} Z_e -1) e \bar{\psi} \gamma^\mu \psi A_\mu.
\end{align}
where $\mathcal{L}_{\textrm{R,QED}}$ contains only renormalized quantities and $\mathcal{L}_{\textrm{CT}}$ provides the counter terms. The renormalization constants are defined as
\begin{equation}
e_0 = Z_e e, \quad m_0 = Z_m m, \quad \xi_0 = Z_\xi \xi, \quad
\psi_0 = Z^{1/2}_\psi \psi, \quad A_{0,\mu} = Z_A^{1/2} A_\mu,
\end{equation}
where the subscript 0 denotes bare quantities. From the Ward identities for the photon propagator and for the electron-photon vertex we obtain that
\begin{equation}
Z_\xi = Z_A, \quad Z_e = 1/\sqrt{Z_A}.
\end{equation}
Our task is, therefore, to determine $Z_A$, $Z_\psi$ and $Z_m$ at 1-loop in three different schemes.

In the minimal subtraction schemes we demand that the two-point functions of the electron and the photon are finite at 1-loop, i.e.
\begin{align}
 i \Gamma^{\psi \bar{\psi}}_R (p,-p) &=
\begin{minipage}[b]{0.1\linewidth}\parbox{15mm}{
\includegraphics{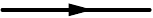}
}
\end{minipage}
\enspace +
\begin{minipage}[b]{0.1\linewidth}\parbox{15mm}{
\includegraphics{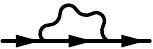}
}
\end{minipage}
\enspace +
\begin{minipage}[b]{0.1\linewidth}\parbox{15mm}{
\includegraphics{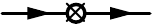}
}
\end{minipage} \enspace  + \mathcal{O}(\alpha^2)  \overset{!}{=} \textrm{finite}
\\
 \leftrightarrow \Gamma^{\psi \bar{\psi}}_R (p,-p) & = (\slashed{p}-m) +
 \Sigma( \slashed{p}) + (Z^{\textrm{MS}/\overline{\textrm{MS}}}_\psi -1) \slashed{p} \nonumber \\
 &- (Z^{\textrm{MS}/\overline{\textrm{MS}}}_\psi Z^{\textrm{MS}/\overline{\textrm{MS}}}_m - 1)m + \mathcal{O}(\alpha^2) \overset{!}{=} \textrm{finite} \label{eq:else}
\end{align}
and
\begin{align}
 -i \left (\Gamma^{A A}_R\right )^{\mu \nu} (q,-q) &= \enspace
\begin{minipage}[b]{0.1\linewidth}\parbox{12mm}{
\includegraphics{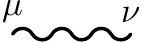}
}
\end{minipage}  + \enspace
\begin{minipage}[b]{0.1\linewidth}\parbox{12mm}{
\includegraphics{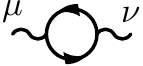}
}
\end{minipage}
 +  \enspace
\begin{minipage}[b]{0.1\linewidth}\parbox{12mm}{
\includegraphics{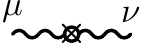}
}
\end{minipage}  \enspace + \mathcal{O}(\alpha^2)\overset{!}{=} \textrm{finite}
\\
\leftrightarrow  \left (\Gamma^{AA}_R\right )^{\mu \nu} (q,-q) & =
(q^2 g^{\mu \nu} - q^\mu q^\nu (1- \xi)) +
\Pi^{\mu \nu} (q) \nonumber \\
& + (q^2 g^{\mu \nu} - q^\mu q^\nu) (Z^{\textrm{MS}/\overline{\textrm{MS}}}_A - 1) + \mathcal{O}(\alpha^2)  \overset{!}{=} \textrm{finite} \label{eq:photse}
\end{align}
Using that
\begin{align}
\begin{minipage}[b]{0.1\linewidth}\parbox{15mm}{
\includegraphics{qed-elprop-2.pdf}
}
\end{minipage} \quad & \equiv  i \Sigma (\slashed{p}) = i (\slashed{p} \Sigma_V (p^2) + m \Sigma_S (p^2)), \\
\begin{minipage}[b]{0.1\linewidth}\parbox{15mm}{
\includegraphics{qed-photprop-2.pdf}
}
\end{minipage}\quad  & \equiv  - i \Pi^{\mu \nu} (q) = - i (q^2 g^{\mu \nu} - q^\mu q^\nu) \Pi (q^2)
\end{align}
and $Z_i = 1 + \delta_i + \mathcal{O}(\alpha^2)$ we can rewrite Eqs.\ \eqref{eq:else} and \eqref{eq:photse} as

\begin{align}
\slashed{p}(\delta^{\textrm{MS}/\overline{\textrm{MS}}}_\psi + \Sigma_V(p^2)) - m (\delta^{\textrm{MS}/\overline{\textrm{MS}}}_\psi + \delta^{\textrm{MS}/\overline{\textrm{MS}}}_m - \Sigma_S(p^2)) + \mathcal{O}(\alpha^2) &\overset{!}{=} \textrm{finite},  \label{eq:mscel} \\
\Pi(q^2) + \delta^{\textrm{MS}/\overline{\textrm{MS}}}_A + \mathcal{O}(\alpha^2) &\overset{!}{=} \textrm{finite} \label{eq:mscphot},
\end{align}
where we dropped the manifestly finite terms $(\slashed{p}-m)$ and $(q^2 g^{\mu \nu} - q^\mu q^\nu (1- \xi))$, since in minimal subtraction schemes we are interested only in subtracting the singularity $1/\eps$ (MS scheme) or $1/\eps - \gamma_E + \log(4\pi)$ ($\overline{\textrm{MS}}$ scheme).

The renormalization condition for the electron-photon vertex reads

\begin{align}
 i e \left( \Gamma^{A \psi \bar{\psi}}_R \right )^\mu (p_1, -p_2) &= \enspace
\begin{minipage}[b]{0.1\linewidth}\parbox{15mm}{
\includegraphics{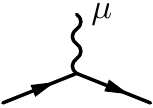}
}
\end{minipage}
 \enspace + \enspace
\begin{minipage}[b]{0.1\linewidth}\parbox{15mm}{
\includegraphics{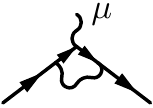}
}
\end{minipage}
 \enspace + \enspace
\begin{minipage}[b]{0.1\linewidth}\parbox{15mm}{
\includegraphics{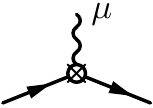}
}
\end{minipage} \enspace + \mathcal{O}(\alpha^2) \overset{!}{=} \textrm{finite}
\\
\leftrightarrow \left( \Gamma^{A \psi \bar{\psi}}_R \right )^\mu (p_1, -p_2)& =
\gamma^\mu + \mathcal{V}^\mu(p_1,                                                    p_2) \nonumber \\
&+ \left (Z^{\textrm{MS}/\overline{\textrm{MS}}}_\psi \sqrt{Z^{\textrm{MS}/\overline{\textrm{MS}}}_A} Z^{\textrm{MS}/\overline{\textrm{MS}}}_e -1 \right) \gamma^\mu  + \mathcal{O}(\alpha^2) \overset{!}{=} \textrm{finite}  \label{eq:vertex},
\end{align}
where
\begin{equation}
 \begin{minipage}[b]{0.1\linewidth}\parbox{15mm}{
\includegraphics{qed-vertex-2.pdf}
}
\end{minipage} \quad \equiv i e \mathcal{V}^\mu(p_1,p_2) .
\end{equation}
The condition given in Eq.\ \eqref{eq:vertex} is equivalent to
\begin{equation}
\mathcal{V}^\mu(p_1,p_2) + \left (\frac{1}{2} \delta^{\textrm{MS}/\overline{\textrm{MS}}}_A + \delta^{\textrm{MS}/\overline{\textrm{MS}}}_e + \delta^{\textrm{MS}/\overline{\textrm{MS}}}_\psi \right ) \gamma^\mu \overset{!}{=} \textrm{finite}.
\end{equation}
Since $Z_e$ has already been fixed by the Ward's identity, it is not necessary to explicitly evaluate the vertex function. On the other hand, nothing prevents us from doing so as a cross-check for the whole calculation.

In the on-shell scheme one demands that the renormalized self-energies satisfy
 \begin{align}
  \lim_{p^2 \to m^2} \left ( \frac{1}{\slashed{p}-m} \Gamma^{\psi \bar{\psi}}_R (p,-p) u(p) \overset{!}{=} u(p) \right ), \\
 \lim_{q^2 \to 0} \left ( \frac{\left (\Gamma^{AA}_R\right )^{\mu \nu} (q,-q)}{q^2} \varepsilon_\nu (q) \overset{!}{=} - \varepsilon^\mu (q) \right ).
 \end{align}
It is easy to show that this corresponds to the following conditions
\begin{align}
 \delta^{\textrm{OS}}_m & \overset{!}{=} \Sigma_S(m^2) + \Sigma_V(m^2), \label{eq:osdelm}  \\
\delta^{\textrm{OS}}_\psi &\overset{!}{=} - \Sigma_V(m^2) + 2m^2 ( \Sigma'_S(m^2) + \Sigma'_V(m^2)), \label{eq:osdelpsi} \\
\delta^{\textrm{OS}}_A &\overset{!}{=} - \lim_{q^2 \to 0}  \frac{\partial}{\partial q^2} \left( q^2 \Pi(q^2) \right ). \label{eq:osdela}
\end{align}
The on-shell condition for the vertex is given by
\begin{equation}
\bar{u} (p)  \left( \Gamma^{A \psi \bar{\psi}}_R \right )^\mu (p, -p) u(p) =  \bar{u} (p)  \gamma^\mu u(p),
\end{equation}
with $p^2 = m^2$. It should also be remarked that, apart from the renormalization, the evaluation of the on-shell vertex function allows us to extract some beautiful piece of physics. Owing to Lorentz invariance, the
 vertex function sandwiched between two spinors with $p_1 \neq p_2$ can be parametrized as
\begin{equation}
\bar{u} (p_2) \left( \Gamma^{A \psi \bar{\psi}}_R \right )^\mu (p_1, -p_2) u (p_1) \overset{!}{=} \bar{u} (p_2) \left ( \gamma^\mu F_1 (q^2) + \frac{i \sigma^{\mu \nu} q_\nu}{2 m} F_2 (q^2) \right ) u (p_1) \label{eq:vertexfu},
\end{equation}
with $q \equiv p_2 - p_1$ and $ \sigma^{\mu \nu} = \frac{i}{2}[\gamma^\mu,\gamma^\nu]$. Here, $F_2 (0)$ is the 1-loop quantum correction to the anomalous magnetic moment of the electron in QED  that was first obtained by Schwinger \cite{Schwinger1948} in 1948
\begin{equation}
\frac{g - 2 }{2} = F_2 (0) = \frac{\alpha}{2 \pi}
\end{equation}
and is often considered to be one of the greatest triumphs of QFT.

For $p_2 = p_1 \equiv p$ the term proportional to $F_2(0)$ vanishes and we are left with $F_1(0)$ only, i.e.
\begin{align}
& \bar{u} (p) \left( \Gamma^{A \psi \bar{\psi}}_R \right )^\mu (p, -p) u (p)  = \nonumber \\
& \bar{u} (p) \gamma^\mu u (p) + \bar{u} (p) \mathcal{V}^\mu u (p)
  + \left ( \frac{1}{2} \delta^{\textrm{OS}}_A +  \delta^{\textrm{OS}}_e + \delta^{\textrm{OS}}_\psi  \right ) \bar{u} (p) \gamma^\mu u (p) = \nonumber  \\
& = F_1 (0) \bar{u} (p)  \gamma^\mu u (p) \overset{!}{=}   \bar{u} (p) \gamma^\mu u (p) .
\label{eq:osvert}
\end{align}
This leads to the renormalization condition
\begin{equation}
\bar{u} (p) \mathcal{V}^\mu(p,p) u (p) = - \left ( \frac{1}{2} \delta^{\textrm{OS}}_A +
 \delta^{\textrm{OS}}_e + \delta^{\textrm{OS}}_\psi  \right ). \label{eq:osdele}
\end{equation}

Now that we fully understand what we want to compute, it is time to carry out the calculation. When doing  things by pen and paper, this is usually the most dull part, which however requires great care. With \feyncalc and \feynhelpers one, of course, still has to pay attention to what one is doing, but the calculation itself can be done much faster.

Although \feynarts already contains the full Standard Model Lagrangian including the counter-terms, for this example we choose  to create a model file that contains only QED. This can be conveniently done via \feynrules \cite{Christensen2008}, a \mma package that generates Feynman rules out of the given Lagrangian. The reason for doing so is to show how \feynrules can be chained with \feynarts, \feyncalc and \feynhelpers to generate new models and study their phenomenology, including the
determination of the renormalization coefficients in different schemes. The results can be also used as a cross-check for \nloct \cite{Degrande2014}, a \feynrules extension for fully automatic 1-loop renormalization of arbitrary models.

The \feynrules-model file \texttt{QED.fr} is already included in \feyncalc 9.2 and can be found in \texttt{FeynCalc/Examples/FeynRules/QED}. Evaluating \texttt{GenerateModelQED.m} converts the input file \texttt{QED.fr} into a working \feynarts model and saves it to \texttt{FeynCalc/FeynArts/Models/QED}. Then we start a new \mma kernel and load \feyncalc, \feynarts and \feynhelpers in the usual way

\begin{mmaCell}[moredefined={$LoadAddOns, $LoadFeynArts, $FAVerbose, FeynCalc}]{Input}
  $LoadAddOns=\{"FeynHelpers"\};
  $LoadFeynArts= True;
  <<FeynCalc`
  $FAVerbose = 0;
\end{mmaCell}
First of all we need to patch the new \feynarts QED model in order to make it compatible with \feyncalc. Since in this calculation we will explicitly distinguish between UV and IR divergences regulated by $\eps_\textrm{UV}$ and $\eps_\textrm{IR}$, we also need to activate the corresponding option

\begin{mmaCell}[moredefined={FAPatch, PatchModelsOnly, $KeepLogDivergentScalelessIntegrals}]{Input}
  FAPatch[PatchModelsOnly->True];
  $KeepLogDivergentScalelessIntegrals=True;
\end{mmaCell}

\begin{mmaCell}{Print}
  Patching FeynArts models... done!
\end{mmaCell}
From our QED model we can generate the required 1-loop diagrams and the corresponding counter-terms with \feynarts and then evaluate them with \feyncalc and \feynhelpers. The conversion of a \feynarts amplitude into \feyncalc notation can be conveniently done with \texttt{FCFAConvert}, where we explicitly keep the gauge parameter $\xi$. Our starting point is the electron self-energy $i\Sigma(\slashed{p})$. Tensor reduction with \texttt{TID} allows us to express this amplitude in terms of the Passarino--Veltman functions $A_0$, $B_0$ and $C_0$, while $\texttt{PaXEvaluateUVIRSplit}$ provides the full analytic result for $i\Sigma(\slashed{p})$.

According to Eq.\ \eqref{eq:mscel} we need to determine $\delta_\psi$ and $\delta_m$ in such a way, that the singularity is subtracted. This can be achieved by adding the amplitudes for the 1-loop self-energy and the counter-term and discarding all the finite terms, which gives

\begin{mmaCell}{Output}
  -i\mmaSub{\(\delta\)}{\(\psi\)}\mmaSub{m}{e}-i\mmaSub{m}{e}\mmaSub{\(\delta\)}{m}+\mmaFrac{i\(\alpha\)\mmaSub{\(\Delta\)}{UV}\big(-\mmaSub{m}{e}\mmaSub{\(\xi\)}{V(1)}-3 \mmaSub{m}{e}+\mmaSub{\(\xi\)}{V(1)}\(\gamma\cdot\)p\big)}{4\(\pi\)}+i\mmaSub{\(\delta\)}{\(\psi\)}\(\gamma\cdot\)p
\end{mmaCell}
Equating this to zero and solving for $\delta_\psi$ and $\delta_m$ we obtain
\begin{mmaCell}{Output}
  \{\mmaSubSup{\(\delta\)}{\(\psi\)}{MS}\(\to\) -\mmaFrac{\(\alpha\)\mmaSub{\(\xi\)}{V(1)}}{4\(\pi\)\mmaSub{\(\varepsilon\)}{UV}}, \mmaSubSup{\(\delta\)}{m}{MS}\(\to\) -\mmaFrac{3\(\alpha\)}{4\(\pi\)\mmaSub{\(\varepsilon\)}{UV}}\}
\end{mmaCell}
\begin{mmaCell}{Output}
  \{\mmaSubSup{\(\delta\)}{\(\psi\)}{\mmaOver{MS}{---}}\(\to\) -\mmaFrac{\(\alpha\)\mmaSub{\(\Delta\)}{UV}\mmaSub{\(\xi\)}{V(1)}}{4\(\pi\)}, \mmaSubSup{\(\delta\)}{m}{\mmaOver{MS}{---}}\(\to\) -\mmaFrac{3\(\alpha\)\mmaSub{\(\Delta\)}{UV}}{4\(\pi\)}\}
\end{mmaCell}

In order to compute these renormalization constants in the on-shell scheme we need to extract $\Sigma_S (p^2)$ and $\Sigma_V (p^2)$ first. Eq.\ \ref{eq:osdelm} tells us that the mass renormalization constant follows from the sum of $\Sigma_S$ and $\Sigma_V $ evaluated at $p^2 = m^2$. To compute this integral we expand $\Sigma_S(p^2) + \Sigma_V(p^2)$ around $p^2 = m^2$ to zeroth order with \texttt{PaXSeries} and obtain
\begin{mmaCell}{Output}
  \{\mmaSubSup{\(\delta\)}{m}{OS}\(\to\) -\mmaFrac{\(\alpha\)\big(3\,log\big(\mmaFrac{\mmaSup{\(\mu\)}{2}}{\mmaSubSup{m}{e}{2}}\big)+3\mmaSub{\(\Delta\)}{UV}+4\big)}{4\(\pi\)}\}
\end{mmaCell}
In the same manner we can also compute the values of $\Sigma'_S(m^2) + \Sigma'_V(m^2)$ and $\Sigma_V(m^2)$. Plugging them into Eq. \eqref{eq:osdelpsi} yields
\begin{mmaCell}{Output}
  \{\mmaSubSup{\(\delta\)}{\(\psi\)}{OS}\(\to\) -\mmaFrac{\(\alpha\)\big(3\,log\big(\mmaFrac{\mmaSup{\(\mu\)}{2}}{\mmaSubSup{m}{e}{2}}\big)-\mmaSub{\(\Delta\)}{IR}(\mmaSub{\(\xi\)}{V(1)}-3)+\mmaSub{\(\Delta\)}{UV}\mmaSub{\(\xi\)}{V(1)}+4\big)}{4\(\pi\)}\}
\end{mmaCell}
As is familiar from the literature \cite{Broadhurst1991}, for $\Delta_\textrm{UV} = \Delta_\textrm{IR} = \Delta$ the parameter $\xi$ drops out, leaving us with a gauge independent fermion wave-function renormalization constant  $\delta^\textrm{OS}_\psi = \delta^\textrm{OS}_m$.

The evaluation of the photon self-energy proceeds similarly. From Eq.\ \eqref{eq:mscphot} we get the renormalization constant $\delta_A$ in the minimal subtraction schemes
\begin{mmaCell}{Output}
  \{\mmaSubSup{\(\delta\)}{A}{MS}\(\to\) -\mmaFrac{\(\alpha\)}{3\(\pi\)\mmaSub{\(\varepsilon\)}{UV}}\}
\end{mmaCell}
\begin{mmaCell}{Output}
  \{\mmaSubSup{\(\delta\)}{A}{\mmaOver{MS}{---}}\(\to\) -\mmaFrac{\(\alpha\)\mmaSub{\(\Delta\)}{UV}}{3\(\pi\)}\}
\end{mmaCell}
Then we again use \texttt{PaXSeries} to compute $\lim_{q^2 \to 0}  \frac{\partial}{\partial q^2} \left( q^2 \Pi(q^2) \right )$ and thus determine $\delta_A$ in the on-shell scheme
\begin{mmaCell}{Output}
  \{\mmaSubSup{\(\delta\)}{A}{OS}\(\to\) -\mmaFrac{\(\alpha\)\big(log\big(\mmaFrac{\mmaSup{\(\mu\)}{2}}{\mmaSubSup{m}{e}{2}}\big)+\mmaSub{\(\Delta\)}{UV}\big)}{3 \(\pi\)}\}
\end{mmaCell}

Let us now also do some calculations with the vertex function. Although \texttt{TID} can of course reduce it into basic scalar integrals, the resulting expression will be huge, as we keep the kinematics completely general. We can, however, obtain a much more compact expression by sticking to coefficient functions, i.e.\ by activating the option \texttt{UsePaVeBasis}. Furthermore, as we are interested only in the UV-part of the whole expression, it is sufficient (and also much faster) to use \texttt{PaXEvaluateUV} instead of \texttt{PaXEvaluateUVIRSplit}. Then the UV-part of the vertex diagram reads
\begin{mmaCell}{Output}
  \mmaFrac{i\(\alpha\)e\mmaSup{\(\gamma\)}{Lor1}\mmaSub{\(\xi\)}{V(1)}}{4\(\pi\)\mmaSub{\(\varepsilon\)}{UV}}
\end{mmaCell}
and with Eq.\ \eqref{eq:vertex} we arrive to
\begin{mmaCell}{Output}
  -\mmaFrac{2\(\pi\)\mmaSub{\(\varepsilon\)}{UV}(\mmaSub{\(\delta\)}{A}+2(\mmaSub{\(\delta\)}{\(\psi\)}+\mmaSub{\(\delta\)}{e}))+\(\alpha\)\mmaSub{\(\xi\)}{V(1)}}{4 \(\pi\) \mmaSub{\(\varepsilon\)}{UV}} = 0
\end{mmaCell}
Plugging our results for $\delta_\psi$ and $\delta_m$ into this relation yields
\begin{mmaCell}{Output}
  \mmaSub{\(\delta\)}{A}+2\mmaSub{\(\delta\)}{e} = 0
\end{mmaCell}
which explicitly confirms the relation $Z_e = 1/\sqrt{Z_A}$ in the minimal subtraction schemes.

To confirm this relation also in the on-shell scheme we need to look at the  vertex function sandwiched between two spinors with their 4-momenta being put on-shell. At first, we choose the momenta to be $p_1$ and $p_2$ with $p_1 \neq p_2$. After having applied Gordon decomposition
\begin{equation}
\bar{u}(p_2) \gamma^\mu u(p_1) = \bar{u}(p_2) \left( \frac{(p_1+p_2)^\mu}{2m} + \frac{i \sigma^{\mu \nu} (p_{2} - p_{1})_\nu}{2m} \right ) u(p)
\end{equation}
we can bring the amplitude to the form dictated by  Eq.\ \eqref{eq:vertexfu}. As neither the counter-term nor the tree-level vertex contain a term proportional to $\sigma^{\mu \nu}$, the 1-loop vertex amplitude alone is sufficient to extract $F_2(q^2)$
\begin{mmaCell}{Output}
  16\mmaSup{\(\pi\)}{3}\(\alpha\)\mmaSubSup{m}{e}{2}\big(2\mmaSub{C}{1}(\mmaSubSup{m}{e}{2}, 2\mmaSubSup{m}{e}{2}-2(p1\(\cdot\)p2), \mmaSubSup{m}{e}{2}, 0, \mmaSubSup{m}{e}{2}, \mmaSubSup{m}{e}{2})+(D-2)\big(\mmaSub{C}{11}(\mmaSubSup{m}{e}{2}, 2\mmaSubSup{m}{e}{2}-2(p1\(\cdot\)p2), \mmaSubSup{m}{e}{2}, 0, \mmaSubSup{m}{e}{2}, \mmaSubSup{m}{e}{2})+\mmaSub{C}{12}(\mmaSubSup{m}{e}{2}, 2\mmaSubSup{m}{e}{2}-2(p1\(\cdot\)p2), \mmaSubSup{m}{e}{2}, 0, \mmaSubSup{m}{e}{2}, \mmaSubSup{m}{e}{2})\big)\big)
\end{mmaCell}
Evaluating it at $q^2=0$ we recover the famous result for $F_2 (0) = \frac{g-2}{2}$
\begin{mmaCell}[moredefined={PaXEvaluateUVIRSplit, tmp, FCI, SPD, SMP, PaXImplicitPrefactor}]{Input}
  PaXEvaluateUVIRSplit[tmp[15]/.\{FCI[SPD[p1,p2]]\(\pmb{\to}\)SMP["m_e"]^2\},PaXImplicitPrefactor\(\pmb{\to}\)1/(2Pi)^D]
\end{mmaCell}
\begin{mmaCell}{Output}
  \mmaFrac{\(\alpha\)}{2\(\pi\)}
\end{mmaCell}
When we set $p_1 = p_2 \equiv p$, the term proportional to $F_2(0)$ vanishes. Extracting $F_1(0)$ from $\bar{u} (p) \left( \Gamma^{A \psi \bar{\psi}}_R \right )^\mu (p, -p) u (p)$ according to Eq.\ \eqref{eq:osvert} tells us that
\begin{mmaCell}{Output}
  -\mmaFrac{2\(\pi\)\mmaSub{\(\varepsilon\)}{UV}(\mmaSub{\(\delta\)}{A}+2(\mmaSub{\(\delta\)}{\(\psi\)}+\mmaSub{\(\delta\)}{e}))+\(\alpha\)\mmaSub{\(\xi\)}{V(1)}}{4\(\pi\)\mmaSub{\(\varepsilon\)}{UV}} = 0
\end{mmaCell}
and hence again
\begin{mmaCell}{Output}
  \mmaSub{\(\delta\)}{A}+2\mmaSub{\(\delta\)}{e} = 0
\end{mmaCell}
as it should.

This concludes the 1-loop renormalization of QED in 3 different schemes, carried out with \feyncalc, \feynhelpers, \feynarts and \feynrules. Needless to say that similar calculations can also be done for more complicated theories (e.g., new models for physics beyond the Standard Model).
\subsection{Quark-gluon vertex expanded in the relative momentum squared at 1-loop}

The ability of \feynhelpers to distinguish between dimensionally regulated UV and IR divergences at 1-loop and to expand coefficient functions in their parameters can be also used to reproduce parts of the 1-loop matching between QCD and
nonrelativistic QCD (NRQCD) \cite{Bodwin1995}. NRQCD is an EFT that is widely used to describe production and decay of heavy quarkonia. In \cite{Shtabovenko2016} we have already demonstrated how \feyncalc can be employed for the tree-level matching between both theories. Here we concentrate on the 1-loop matching by following the famous work of Manohar \cite{Manohar1997},
where on the QCD side of the matching the on-shell vertex function was evaluated using background field formalism \cite{Abbott1981,Abbott1982} and expanded up to the first order in the relative momentum squared.

As \feynarts does not contain a model for background-field QCD, we chose to implement it ourselves using \feynrules and to include the model file \texttt{QCDBGF.fr} to \feyncalc 9.2\footnote{The model file is located in \texttt{FeynCalc/Examples/FeynRules/QCDBGF}.}.
A working \feynarts model can be generated via \texttt{GenerateModelQCDBGF.m}.

The abelian and nonabelian diagrams can be parametrized as

\begin{align}
\begin{minipage}[b]{0.1\linewidth}\parbox{25mm}{
\includegraphics{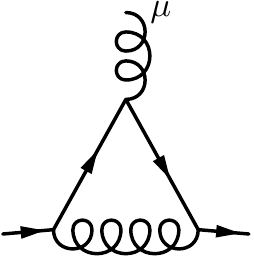}
}
\end{minipage} \quad \quad = - i g T^a \bar{u}(p_2) \left ( F^{(V)}_1 (q^2) \gamma^\mu + i F^{(V)}_2 (q^2) \frac{\sigma^{\mu \nu} q_\nu}{2m} \right ) u(p_1), \\
\nonumber \\
\begin{minipage}[b]{0.1\linewidth}\parbox{25mm}{
\includegraphics{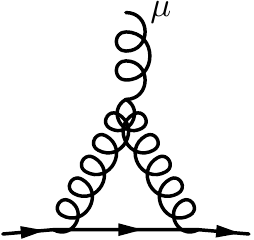}
}
\end{minipage} \quad \quad = - i g T^a \bar{u}(p_2) \left ( F^{(g)}_1 (q^2) \gamma^\mu + i F^{(g)}_2 (q^2) \frac{\sigma^{\mu \nu} q_\nu}{2m} \right ) u(p_1),
\end{align}
where $q \equiv p_2 = p_1$. Our goal is to compute the form-factors $F^{(V)}_{1/2}(q^2)$ and $F^{(g)}_{1/2}(q^2)$ expanded up to $\mathcal{O}(q^2/m^2)$.  In this case the whole calculation can be essentially split into 5 distinct steps
\begin{enumerate}
\item Generate the diagrams with \feynarts and prepare amplitudes for \feyncalc.
\item Perform tensor decomposition of 1-loop integrals and simplify the Dirac algebra.
\item Apply Gordon decomposition.
\item Expand the amplitudes in $q^2$ with \feynhelpers.
\item Extract the form-factors.
\end{enumerate}
Someone familiar with \feyncalc can prepare the corresponding code with minimal effort. Applying it to each of the diagrams we find that $F^{(V)}_1(q^2)$ and $F^{(V)}_2(q^2)$ are given by

\begin{mmaCell}{Output}
  \mmaFrac{1}{\(\pi\)}\mmaSub{\(\alpha\)}{s}\bigg(\mmaSub{C}{F}-\mmaFrac{\mmaSub{C}{A}}{2}\bigg)\bigg(\mmaFrac{\mmaSub{\(\Delta\)}{IR}}{2}-\mmaFrac{q2\mmaSub{\(\Delta\)}{IR}}{6\mmaSup{m}{2}}+\mmaFrac{3}{4}log(\mmaFrac{\mmaSup{\(\mu\)}{2}}{\mmaSup{m}{2}})-\mmaFrac{q2(4\,log(\mmaFrac{\mmaSup{\(\mu\)}{2}}{\mmaSup{m}{2}})+3)}{24\mmaSup{m}{2}}+\mmaFrac{\mmaSub{\(\Delta\)}{UV}}{4}+1\bigg)
\end{mmaCell}
and
\begin{mmaCell}{Output}
  \mmaFrac{\big(\mmaFrac{q2}{12\mmaSup{m}{2}}+\mmaFrac{1}{2}\big)\mmaSub{\(\alpha\)}{s}\big(\mmaSub{C}{F}-\mmaFrac{\mmaSub{C}{A}}{2}\big)}{\(\pi\)}
\end{mmaCell}
respectively, while the nonabelian form-factors $F^{(g)}_1(q^2)$ and $F^{(g)}_2(q^2)$ read
\begin{mmaCell}{Output}
  \mmaFrac{1}{8 \(\pi\)}\mmaSub{C}{A}\mmaSub{\(\alpha\)}{s}\bigg(2\mmaSub{\(\Delta\)}{IR}-\mmaFrac{3\,q2\mmaSub{\(\Delta\)}{IR}}{2\mmaSup{m}{2}}+3\,log\big(\mmaFrac{\mmaSup{\(\mu\)}{2}}{\mmaSup{m}{2}}\big)-\mmaFrac{q2(3\,log\big(\mmaFrac{\mmaSup{\(\mu\)}{2}}{\mmaSup{m}{2}}\big)+2)}{2 \mmaSup{m}{2}}+\mmaSub{\(\Delta\)}{UV}+4\bigg)
\end{mmaCell}
and
\begin{mmaCell}{Output}
  \mmaFrac{1}{8 \(\pi\)}\mmaSub{C}{A}\mmaSub{\(\alpha\)}{s}\bigg(2\mmaSub{\(\Delta\)}{IR}+\mmaFrac{2\,q2\mmaSub{\(\Delta\)}{IR}}{\mmaSup{m}{2}}+2\bigg(log\bigg(\mmaFrac{\mmaSup{\(\mu\)}{2}}{\mmaSup{m}{2}}\bigg)+3\bigg)+\mmaFrac{q2(2\,log\big(\mmaFrac{\mmaSup{\(\mu\)}{2}}{\mmaSup{m}{2}}\big)+1)}{\mmaSup{m}{2}}\bigg)
\end{mmaCell}
where $\texttt{q2} \equiv q^2$. To compare this to the results presented in \cite{Manohar1997}, we need to switch from $D= 4-2 \eps$ to $D= 4-\eps$ via $1/\eps \to 2/\eps$ and eliminate $\gamma_E$ and $\log (4\pi)$ by substituting $\mu^2$ with $\mu^2 \frac{e^{\gamma_E}}{4 \pi}$. After doing so we precisely recover Eqs.\ (24)-(25) and Eqs.\ (29)-(30) from \cite{Manohar1997}.

\subsection{Higgs decay to two gluons}

Let us consider the  partial decay width of the Higgs boson into two gluons via a top-quark loop\footnote{The calculation of this process with \pax alone can be found in the official tutorial of the package, Sec 5.2.}. With \feyncalc, \feynarts and \feynhelpers the calculation is very similar to the two previous examples, the main difference being that for the generation of the diagrams we use \texttt{SM.mod}, the default Standard Model implementation shipped with \feynarts. As the amplitude for this process is finite, it is sufficient to use \texttt{PaXEvaluate} only. Squaring the amplitude, averaging over the polarizations of the on-shell gluons and multiplying by the phase space factor we arrive at the following expression for $\Gamma (H \to gg)$

\begin{mmaCell}{Output}
  \mmaFrac{\mmaSub{G}{F}\mmaSubSup{m}{H}{3}\mmaSubSup{\(\alpha\)}{s}{2}\mmaSup{\bigg((\(\tau\)-1)\mmaSup{log}{2}\bigg(2\bigg(\mmaSqrt{\mmaFrac{\(\tau\)-1}{\(\tau\)}}-1\bigg)\(\tau\)+1\bigg)-4\(\tau\)\bigg)}{2}}{256\mmaSqrt{2}\mmaSup{\(\pi\)}{3}\mmaSup{\(\tau\)}{4}}
\end{mmaCell}
with $\tau = \frac{m_H^2}{4 m_t^2}$, where $m_H$ denotes the Higgs mass, $m_t$ the mass of the top quark and $G_F$ stands for the Fermi constant. According to the literature \cite{SpiraNucl.Phys.B453:17-821995}, $\Gamma (H \to gg)$ (taking into account all quark flavors) can be written as
\begin{equation}
\Gamma (H \to gg) = \frac{G_F \alpha_s^2}{36 \sqrt{2} \pi^3} m_H^3 \biggl | \frac{3}{4} \sum_{\substack{Q=u,d,\\c,s,t,b}} A_Q (\tau_Q) \biggr |^2,
\end{equation}
where $\tau_Q = \frac{m_H^2}{4 m_Q^2}$ and $A_Q$ is defined as
\begin{equation}
A_Q (\tau_Q) = \frac{2}{\tau_Q^2} \left [ \tau_Q + (\tau_Q-1) f(\tau_Q) \right ] \label{eq:aq}
\end{equation}
with
\begin{equation}
f(\tau_Q) = \begin{cases}
\arcsin^2 (\sqrt{\tau_Q}) & \mbox{if } \tau_Q \leq 1 \\
- \frac{1}{4} \left [ \log \left ( \frac{1+\sqrt{1+\tau^{-1}}}{1-\sqrt{1-\tau^{-1}}} \right ) - i \pi \right ]^2 & \mbox{if } \tau_Q > 1 \\
\end{cases}
\end{equation}
Depending on the quark mass, one can have $\tau_Q \leq 1$ or $\tau_Q > 1$, while for the top quark and the known Higgs mass only the former case is relevant. Nevertheless, extracting $A_t^2 (\tau_t) \equiv A^2 (\tau)$ from our result and comparing it to Eq.\ \eqref{eq:aq}
\begin{figure}
  \centering
  \begin{minipage}[b]{0.47\textwidth}
    \includegraphics[width=\textwidth]{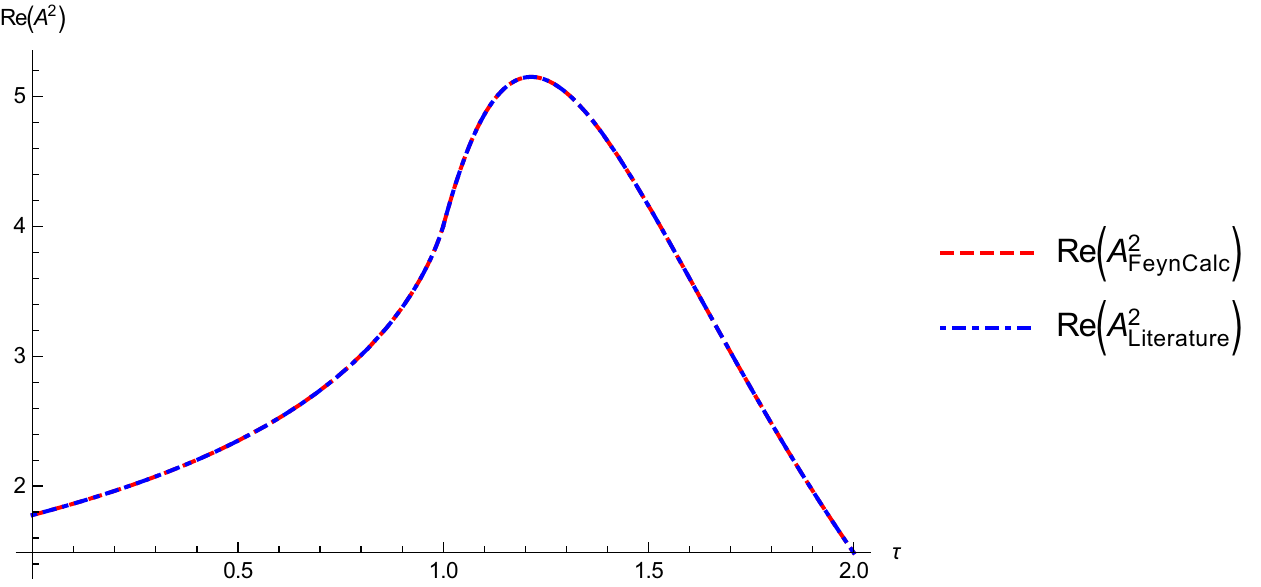}
  \end{minipage}
  \hfill
  \begin{minipage}[b]{0.47\textwidth}
    \includegraphics[width=\textwidth]{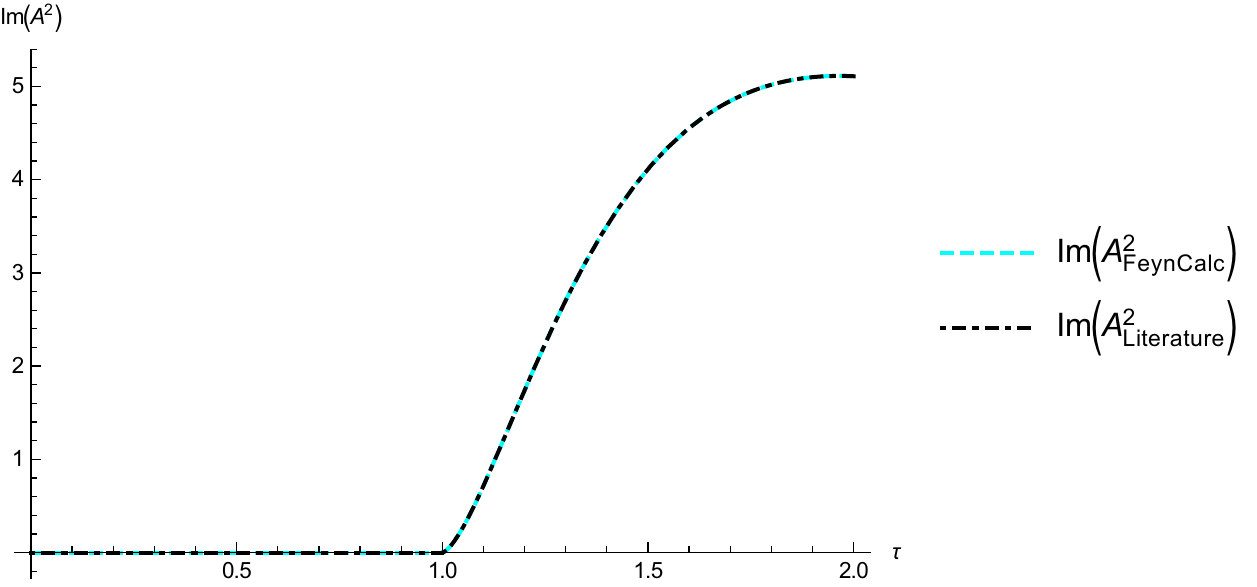}
  \end{minipage}
      \caption{Real and imaginary parts of $A(\tau)$ from the literature and from the result obtained with \feynhelpers.} \label{fig:higgs}
\end{figure}
we can convince ourselves (c.f.\ Fig. \ref{fig:higgs} ) that the analytic expression returned by \pax is indeed valid both for $\tau \leq 1$ and $\tau > 1$, such that our result is correct.

\subsection{2-loop self-energies in massless QED}
Our last example deals with the photon and electron self-energies (with full gauge dependence) in massless QED at 2-loops.
This requires evaluation of six 2-loop diagrams
\begin{align}
\begin{minipage}[b]{0.1\linewidth}\parbox{20mm}{
 \includegraphics{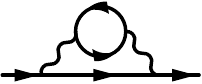}
}
\end{minipage}
\quad \quad +  \quad
\begin{minipage}[b]{0.1\linewidth}\parbox{20mm}{
 \includegraphics{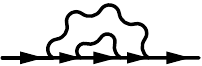}
}
\end{minipage}
\quad \quad +  \quad
\begin{minipage}[b]{0.1\linewidth}\parbox{20mm}{
 \includegraphics{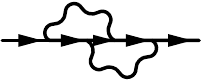}
}
\end{minipage}
\quad \quad  & \equiv  i \slashed{p} \Sigma_{2V} (p^2) \label{eq:se2loopamps}, \\
\begin{minipage}[b]{0.1\linewidth}\parbox{20mm}{
 \includegraphics{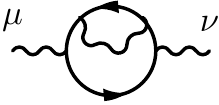}
}
\end{minipage}
\quad \quad + \quad
\begin{minipage}[b]{0.1\linewidth}\parbox{20mm}{
 \includegraphics{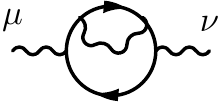}
}
\end{minipage}
\quad \quad + \quad
\begin{minipage}[b]{0.1\linewidth}\parbox{20mm}{
 \includegraphics{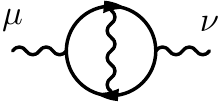}
}
\end{minipage}
 \quad \quad &\equiv  - i (p^2 g^{\mu \nu} - p^\mu p^\nu) \Pi_2 (p^2),
\end{align}
 that can be rewritten in terms of only two master integrals. Here we need to handle the amplitudes in a slightly different way, as compared to the previous examples. In particular, for the decomposition of 2-loop tensor integrals we use \texttt{FCMultiLoopTID} instead of \texttt{TID}, that works only at 1-loop. IBP-reduction with \texttt{FIREBurn} is then used to reduce all the resulting scalar integrals into two master integrals. These main steps can be summarized in only three lines of code for each class of diagrams. For definiteness, let us consider self-energy amplitudes from Eq.\ \eqref{eq:se2loopamps} and denote the expression obtained from \feynarts and processed with \texttt{FCFAConvert} with \texttt{ampsSE}. The sum of the amplitudes depends on the loop momenta $l_1$ and $l_2$ and the external momentum $p$. With
\begin{mmaCell}[moredefined={ampsSE1, ampsSE, DiracTrace, Tr, FCMultiLoopTID, DiracSimplify}]{Input}
  ampsSE1 = (ampsSE /. DiracTrace \(\pmb{\to}\) Tr) //  FCMultiLoopTID[#, \{l1, l2\}] & // DiracSimplify
\end{mmaCell} 
we evaluate Dirac traces, carry out the tensor decomposition and simplify the resulting Dirac structures. Then, in 
\begin{mmaCell}[moredefined={ampsSE1, ampsSE2, FDS, DiracSimplify}]{Input}
  ampsSE2 = ampsSE1 // FDS[#, l1, l2] &
\end{mmaCell} 
we use \texttt{FDS} to simplify loop integrals by shifting their loop momenta.  The IBP-reduction is started by applying \texttt{FIREBurn} to the resulting expression
\begin{mmaCell}[moredefined={ampsSE3, ampsSE2, FDS, FIREBurn}]{Input}
  ampsSE3 = FIREBurn[ampsSE2, \{l1, l2\}, \{p\}] // FDS[#, l1, l2] &
\end{mmaCell} 
Finally, after sorting terms using \feyncalc's \texttt{Collect2} (more advanced version of \mma's \texttt{Collect})
\begin{mmaCell}[moredefined={ampsSE3, ampsSE4, FDS, FIREBurn, Collect2,FeynAmpDenominator,Factoring}]{Input}
  ampsSE4=ampsSE3//Collect2[#,\{FeynAmpDenominator\},Factoring \(\pmb{\to}\) FullSimplify]&
\end{mmaCell} 
and factoring out $i \slashed{p}$
\begin{mmaCell}[moredefined={ampsSE1, ampsSE2,  ampsSE4, resSE, Cancel, FDS, DiracSimplify, FCI,GSD}]{Input}
  resSE=Cancel[ampsSE4/(-I FCI[(GSD[p])])]//FullSimplify
\end{mmaCell} 
we find that the 2-loop contribution\footnote{with $1/(2\pi)^{2D}$ omitted.} to $\Sigma_{2V}(p^2)$ equals
\begin{mmaCell}{Output}
  \mmaFrac{(D-2)\mmaSup{e}{4}(\mmaFrac{2((D-6)(D-3)(3 D-8)\mmaSubSup{\(\xi\)}{A}{2}-D((D-9)D+6)-40)}{\mmaSup{l1}{2}.\mmaSup{(l1-l2)}{2}.\mmaSup{(l2-p)}{2}}-\mmaFrac{(D-6)(D-4)\mmaSup{p}{2}((D-2)\mmaSubSup{\(\xi\)}{A}{2}+D-6)}{\mmaSup{l1}{2}.\mmaSup{l2}{2}.\mmaSup{(l1-p)}{2}.\mmaSup{(l2-p)}{2}})}{4(D-6)(D-4)\mmaSup{p}{2}}
\end{mmaCell}
The treatment of  vacuum polarization diagrams proceeds in the same fashion and yields
\begin{mmaCell}{Output}
  \mmaFrac{2(D-2)\mmaSup{e}{4}(\mmaFrac{4(D-3)((D-4)D+8)}{\mmaSup{l1}{2}.\mmaSup{(l1-l2)}{2}.\mmaSup{(l2-p)}{2}}-\mmaFrac{(D-4)((D-7)D+16)\mmaSup{p}{2}}{\mmaSup{l1}{2}.\mmaSup{l2}{2}.\mmaSup{(l1-p)}{2}.\mmaSup{(l2-p)}{2}})}{\mmaSup{(D-4)}{2}(D-1)\mmaSup{p}{2}}
\end{mmaCell}
for $\Pi_2 (p^2)$. 

As expected, the vacuum polarization amplitude is gauge invariant, while the electron self-energy depends on the gauge parameter $\xi$. These results precisely agree with the literature, e.g. Eq.\ 5.18 and Eq.\ 5.51 from \cite{Grozin2007d}.

\section{Summary}
We have presented the first stable public version of an easy-to-use interface called \feynhelpers that seamlessly integrates the 1-loop library of \pax and  the IBP-reduction mechanism of \fire into \feyncalc. With this add-on many types of calculations that were difficult or hardly feasible with \feyncalc previously, can now be done in a much simpler way. This was demonstrated with four different examples from QED, QCD and Higgs physics. The interface code is open-source and can be modified to accommodate specific requirements. The goals for the future development  of \feynhelpers are to further improve the integration with \pax and \fire but also to add new interfaces to interesting and useful HEP tools, like \litered \cite{Lee2012} or \formtracer\cite{Cyrol2016}.

\section*{Acknowledgments}

The author would like to thank Hiren Patel, the developer of \pax, and Alexander Smirnov, the developer of \fire, for many useful explanations on the usage of their respective packages. He appreciates useful feedback provided by Simone Biondini while using an early version of \feynhelpers to cross-check analytic results for some loop integrals computed in \cite{Biondini2015,Biondini2016}. Special gratitude also goes to Norbert Kaiser for his help in checking some results obtained with \feynhelpers and to Rolf Mertig and Frederik Orellana for careful reading of this manuscript. Nora Brambilla, Antonio Vairo and Andrey Grozin are acknowledged for interesting discussions and insightful comments.

 This work has been supported by the DFG and the NSFC through funds provided to the Sino-German CRC 110 ``Symmetries and the Emergence of Structure in QCD'' (NSFC Grant No. 11261130311), and by the DFG cluster of excellence ``Origin and structure of the universe'' (\url{www.universe-cluster.de}).

\end{document}